\documentclass[floatfix,
aps,
10pt,
twocolumn,
superscriptaddress,
amsmath,amssymb,
physrev,
floatfix,
]{revtex4-2}

\usepackage{amsmath}
\usepackage{amssymb}
\usepackage[]{graphicx}
\usepackage{iftex}
\usepackage[colorlinks=true,citecolor=blue,linktocpage=true,linkcolor=blue,filecolor=blue,urlcolor=blue]{hyperref}
\usepackage{physics}
\usepackage{bbm}
\usepackage{bm}
\usepackage{array}
\usepackage{algorithm}
\usepackage{algpseudocode}
\usepackage{dcolumn}
\usepackage{color}
\usepackage{here}

\ifdefined\annotatedversion
\newcommand{\si}[1]{{\color{magenta}{#1}}}
\newcommand{\sii}[1]{{\color{blue}{#1}}}
\else
\newcommand{\si}[1]{{\color{black}{#1}}}
\newcommand{\sii}[1]{{\color{black}{#1}}}
\fi

\newcommand{\ha}{\hat{a}}
\newcommand{\had}{\hat{a}^\dagger}

\makeatletter
\def\maketitle{
\@author@finish
\title@column\titleblock@produce
\suppressfloats[t]}
\makeatother

\begin{document}
\preprint{APS/123-QED}

\title{
Systematic frequency-collision analysis of the cross-resonance gate \\ outside the straddling regime
}

\author{Shinichi Inoue}
\email{inoue@qipe.t.u-tokyo.ac.jp}
\affiliation{Department of Applied Physics, Graduate School of Engineering, The University of Tokyo, Bunkyo-ku, Tokyo 113-8656, Japan}

\author{Shotaro Shirai}
\affiliation{RIKEN Center for Quantum Computing (RQC), Wako, Saitama 351--0198, Japan}
\affiliation{Komaba Institute for Science (KIS), The University of Tokyo, Meguro-ku, Tokyo, 153-8902, Japan}

\author{Shuhei Tamate}
\affiliation{RIKEN Center for Quantum Computing (RQC), Wako, Saitama 351--0198, Japan}

\author{Shu~Watanabe}
\altaffiliation{Present address: Quantum Laboratory, Fujitsu Limited, Nakahara-ku, Kawasaki, Kanagawa 211-8588, Japan}
\affiliation{Department of Applied Physics, Graduate School of Engineering, The University of Tokyo, Bunkyo-ku, Tokyo 113-8656, Japan}

\author{Kohei~Matsuura}
\affiliation{Department of Applied Physics, Graduate School of Engineering, The University of Tokyo, Bunkyo-ku, Tokyo 113-8656, Japan}

\author{Rui Li}
\affiliation{RIKEN Center for Quantum Computing (RQC), Wako, Saitama 351--0198, Japan}

\author{Yasunobu Nakamura}
\email{yasunobu@ap.t.u-tokyo.ac.jp}
\affiliation{Department of Applied Physics, Graduate School of Engineering, The University of Tokyo, Bunkyo-ku, Tokyo 113-8656, Japan}
\affiliation{RIKEN Center for Quantum Computing (RQC), Wako, Saitama 351--0198, Japan}

\date{\today}

\begin{abstract}
    Frequency crowding remains a major obstacle to scaling fixed-frequency transmon processors. Among the widely used all-microwave two-qubit gates, the cross-resonance (CR) gate is particularly sensitive to qubit-frequency spread because the conventional straddling regime condition constrains assignable qubit frequencies tightly and makes the system susceptible to frequency collisions. \si{Here, we present a systematic framework for frequency-collision analysis and frequency-allocation optimization, focusing on the CR gate outside the straddling regime, which we refer to as the far-detuned regime. We numerically characterize frequency collisions among qubits through systematic parameter sweeps and identify collision-free conditions. We then optimize frequency allocation using mixed-integer programming on a unit-cell lattice with periodic boundary conditions.} We demonstrate that far-detuned designs significantly reduce collisions compared with designs in the straddling regime. Monte Carlo yield analysis indicates that a 10\% collision-free yield for a 1024-qubit square lattice at \sii{a 0.1\% two-qubit-gate error threshold requires $\sigma_{\mathrm{f}}/2\pi \le 6.86~\mathrm{MHz}$}. \si{Our frequency-allocation optimization framework could also be useful for other two-qubit gates based on fixed-frequency qubits.}
\end{abstract}

\maketitle

\section{Introduction} \label{sec:intro}
    The superconducting qubit platform has advanced rapidly in recent years. Much of the progress has been enabled by the transmon qubit~\cite{koch2007charge,schreier2008suppressing}. Because of its simple circuit and exponential resilience against charge noise, the transmon has become the leading candidate for large-scale integration. State-of-the-art transmon-based devices have achieved single-qubit-gate infidelity below $0.01\%$~\cite{li2023error, chiaro2025active} and two-qubit-gate and readout infidelities below $0.1\%$~\cite{li2024realization, spring2025fast, marxer2025above, inoue2026parametrically}, surpassing the error-correction thresholds set by the surface-code architecture. Further scaling up the number of qubits while maintaining high-fidelity operations requires detailed frequency assignment to avoid frequency collisions that can degrade gate fidelity~\cite{brink2018device, hertzberg2021laser, zhang2022high}.

    Traditionally, the cross-resonance (CR) gate~\cite{rigetti2005protocol, paraoanu2006microwave, chow2011simple,chow2012universal,corcoles2013process, sheldon2016procedure, takita2017experimental} has been adopted as a two-qubit gate between fixed-frequency transmon qubits. It realizes a $ZX_{\pi/2}$ rotation by driving a control qubit at the frequency of a target qubit capacitively coupled to it. To maximize gate speed and suppress residual $ZZ$ interactions, prior designs have preferentially operated in the straddling regime~\cite{sheldon2016procedure,mckay2019three,tripathi2019operation,sundaresan2020reducing,kandala2021demonstration}, where neighboring target-qubit frequencies lie between the $\ket{0}\!\Leftrightarrow\!\ket{1}$ and $\ket{1}\!\Leftrightarrow\!\ket{2}$ transitions of the control qubit. As the system size grows, this requirement tightly constrains assignable frequencies and leads to serious frequency collisions. Several studies~\cite{brink2018device, hertzberg2021laser,zhang2022high} suggest that, at current levels of qubit-frequency spread, the collision-free yield for nearly thousand-qubit layouts collapses, making straightforward scaling challenging.
    
    A key step toward minimizing frequency collisions is to clarify the underlying collision processes that can degrade gate fidelity. Prior works studied these processes using effective-Hamiltonian methods~\cite{magesan2020effective,malekakhlagh2020first,hertzberg2021laser, osman2023mitigation}, Floquet-Hamiltonian methods~\cite{di2022extensible, heya2024floquet, ma2025analysis}, and numerical approaches~\cite{di2022extensible, zhao2022realization, khezri2023measurement, dumas2024measurement, xie2022suppressing, klimov2024optimizing}. They identified parameter conditions that lead to leakage or induce gate errors within the computational subspace. Such collision analyses motivate frequency-allocation optimization strategies that improve tolerance to qubit-frequency spread due to fabrication inhomogeneities~\cite{zhang2025efficient, morvan2022optimizing, mckinney2024spectator, ai2025scalable}. It is also important to develop error-correcting schemes on a qubit lattice with reduced connectivity to avoid frequency collisions. For example, hexagonal and heavy-hexagonal lattices~\cite{mcewen2023relaxing,eickbusch2024demonstrating,chamberland2020topological} can implement a variant of the surface code with modest overhead while significantly alleviating frequency collisions compared with nearest-neighbor square-lattice layouts. In addition, chiplet architectures, in which multiple small-scale collision-free tiles are interconnected, offer a modular route to large-scale processors~\cite{smith2022scaling}.

    Another approach to mitigate frequency collisions between fixed-frequency qubits is to implement two-qubit gates outside the straddling regime. Experimentally, such operations have been demonstrated with direct coupling using the $\ket{20}\!\Leftrightarrow\!\ket{01}$ transition~\cite{krinner2020demonstration}, a transmon coupler~\cite{shirai2023all, shirai2025high}, and a resonator coupler~\cite{chow2013microwave, paik2016experimental, premaratne2019implementation}. However, these approaches come with increased circuit and control-wire complexities and a larger footprint, all of which can limit scalability. Therefore, from the perspective of simplicity and minimal wiring overhead, a design consisting of two fixed-frequency transmon qubits with direct capacitive coupling is preferable.

    In this work, we revisit the CR gate and analyze frequency collisions outside the straddling regime. We design the CR gate in a regime that we call the far-detuned regime, where the control-qubit $\ket{0}\!\Leftrightarrow\!\ket{1}$ transition frequency lies below the $\ket{1}\!\Leftrightarrow\!\ket{2}$ transition frequency of the target qubits. In this regime, the designable qubit-qubit detuning is no longer limited to a narrow parameter regime, enabling more flexible frequency allocation. However, the effective $ZX$ rate becomes smaller in this regime, requiring stronger microwave drives for high-fidelity operations. Such high-power driving induces leakage and unwanted state transitions, especially during the rise and fall times of the pulse. Experimental work has shown that adding smooth ramps to the beginning and end of the pulse can strongly suppress these non-adiabatic transitions even at high drive power~\cite{motzoi2013improving,tripathi2019operation, werninghaus2021leakage, li2024experimental,hyyppa2024reducing, chiaro2025active}. However, these observations have not been applied to large-scale frequency-collision analyses.

    \si{To evaluate the frequency-collision of the far-detuned CR gate, we present a systematic framework for frequency-allocation optimization. We numerically characterize frequency collisions among qubits through systematic parameter sweeps that account for smooth pulse ramps and identify collision-free conditions at a 0.1\% two-qubit-gate error threshold. We then optimize frequency allocation using mixed-integer programming on unit-cell lattices with periodic boundary conditions for square, hexagonal, and heavy-hexagonal architectures. Finally, we evaluate} the device yield of 1024-qubit layouts under realistic qubit-frequency spread.

    This paper is organized as follows. In Sec.~\ref{sec:overview}, we provide an overview of our frequency-collision analysis. In Sec.~\ref{sec:CR_far_detuned}, we review the CR gate mechanism and derive the effective $ZX$ rate in the far-detuned regime. In Sec.~\ref{sec:lta_theory}, we present a numerical collision simulation method that incorporates smooth pulse ramps. In Sec.~\ref{sec:collision}, we analyze frequency collisions based on numerical simulation for both far-detuned and straddling CR gates. In Sec.~\ref{sec:yield}, we perform frequency allocation on various lattices and evaluate the yield under realistic qubit-frequency spread. In Sec.~\ref{sec:discussion}, we summarize our findings and discuss future directions.

\section{Overview} \label{sec:overview}
\begin{figure*}
    \centering
    \includegraphics[]{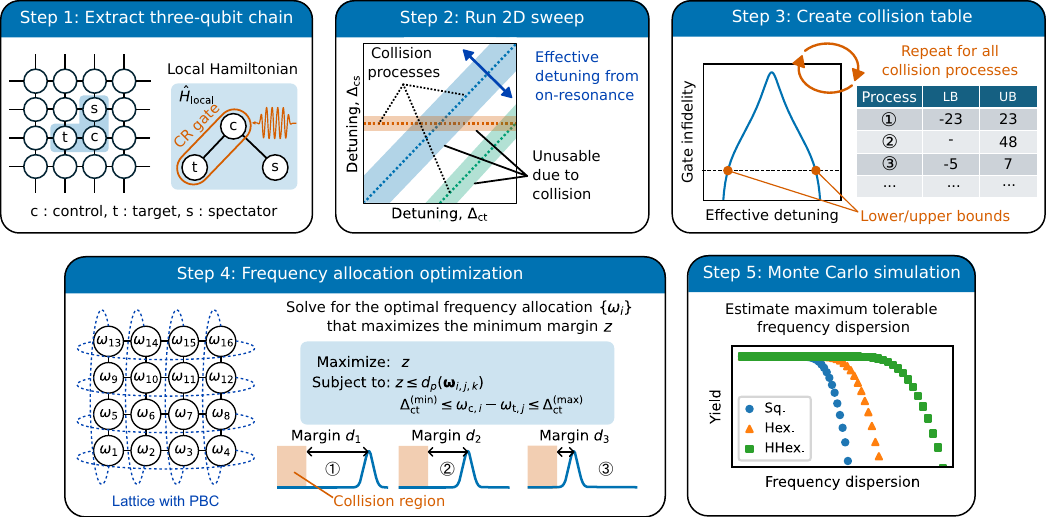}
    \caption{Schematic workflow of the systematic frequency-collision analysis for designing large-scale fixed-frequency transmon processors operated with cross-resonance (CR) gates.
    Step~1:~Extract a local three-qubit chain from the lattice and describe the chain by a local Hamiltonian, where c, t, and s denote the control, target, and spectator qubits, respectively.
    Step~2:~Compute the CR-gate error over a two-dimensional sweep of the relevant qubit--qubit detunings. The systematic parameter sweep generates a frequency-collision landscape in which collisions appear as band-like high-error regions.
    Step~3:~Associate each collision band with an on-resonance condition defined by its effective detuning, and extract the bounds of the effective-detuning window within which the total gate infidelity exceeds a threshold.
    Step~4:~Find optimal qubit frequencies by solving a max--min mixed-integer program on a unit cell with periodic boundary conditions (PBCs) that maximizes the minimum standardized margin over all collision windows.
    Step~5:~Estimate the collision-free chip yield by Monte Carlo sampling of fabrication-induced qubit-frequency spread around the optimized frequency allocation and extract the qubit-frequency spread required to reach 10\% yield.}
    \label{fig:workflow}
\end{figure*}
    This section provides an overview of the frequency-collision analysis studied in this work (Fig.~\ref{fig:workflow}). Our goal is to determine optimal qubit-frequency allocations on finite square, hexagonal, and heavy-hexagonal lattices with periodic boundary conditions. The optimization is performed to maximize the collision-free chip yield. Under this setting, the analysis consists of five steps, which are summarized below.
    
    \vspace{0.5em}
    \textbf{Step 1: Extract a three-qubit chain and define a subsystem Hamiltonian.}
    From the full lattice, we extract a three-qubit chain to analyze frequency collisions. Each chain is modeled by a subsystem Hamiltonian with three transmon qubits: the control c, the target t, and a spectator s that is not directly driven by the CR gate but may be unintentionally excited. There are two chain types depending on which qubit the spectator is coupled to. One is the c-t-s chain~(spectator coupled to the target) and the other is the t-c-s chain~(spectator coupled to the control). For the lattice definition and the extraction of the subsystem Hamiltonian, see Sec.~\ref{sec:collision}~A.

    \vspace{0.5em}
    \textbf{Step 2: Calculate a collision landscape through a two-dimensional detuning sweep.}
    An extracted three-qubit chain is characterized by three qubit frequencies, with one qubit frequency fixed without loss of generality because frequency collisions depend only on relative detunings. We then sweep the two remaining detuning parameters on a two-dimensional grid and compute the CR-drive-induced gate error to obtain a collision landscape. Frequency collision events show band-like high-error regions in this map. This landscape provides a guideline for large-scale quantum processing unit~(QPU) frequency allocation. By avoiding these high-error bands, we keep the gate error below a chosen threshold. See Sec.~\ref{sec:lta_theory}, Sec.~\ref{sec:collision}~B, and Fig.~\ref{fig:collmap_riken} for details.

    \vspace{0.5em}
    \textbf{Step 3: Convert collision landscapes into a collision table and map.}
    Starting from the 2D numerical collision landscape, we first assign each band-like structure to a distinct collision process defined by an effective resonance condition. For example, if a band appears near the line $\Delta_{\mathrm{ct}} = 0$, this process can be interpreted as the frequency collision between the $\ket{0}\!\Leftrightarrow\!\ket{1}$ transitions of the control and target qubits.
    
    For each collision process, we compute the effective detuning between each grid point and the corresponding resonance condition. Plotting the single-peaked state-transition amplitude and the corresponding gate infidelity as a function of this effective detuning around the zero-detuning regions, allows us to extract a collision window: the lower and upper detuning bounds within which the gate error exceeds the chosen infidelity threshold. This process is repeated for all observed bands to create a table and map of collision processes that are relevant for frequency allocation. See Sec.~\ref{sec:collision}~C for details.

    \vspace{0.5em}
    \textbf{Step 4: Optimize frequency allocation using the collision table.} 
    Using the frequency-collision tables from the previous step, we formulate a frequency-allocation problem on a unit cell with periodic boundary conditions (PBCs) so that the optimized solution can be tiled to generate larger layouts. For every three-qubit chain in the unit cell, we evaluate the standardized distance to each collision window and maximize the minimum distance in a \sii{mixed-integer} program. We solve this max--min optimization problem that maximizes the minimum standardized margin over all collision processes and all three-qubit chains, yielding optimal design frequencies that are most tolerant to fabrication-induced qubit-frequency spread. See Sec.~\ref{sec:yield}~A for details.

    \vspace{0.5em}
    \textbf{Step 5: Estimate large-scale yield by Monte Carlo sampling.} 
    Based on the optimized frequency allocations $\qty{\omega_{i}}$, we simulate fabrication-induced qubit-frequency spread by adding independent random frequency shifts and sample many chip realizations. A sample is declared colliding if any three-qubit chain violates the collision-table constraints. The chip yield is the fraction of collision-free samples over all samples. From the yield curve, we extract the qubit-frequency spread required to reach a 10\% device yield for 1024-qubit devices with square, hexagonal, and heavy-hexagonal lattices. See Sec.~\ref{sec:yield}~B for details.

\section{CR gate in the far-detuned regime} \label{sec:CR_far_detuned}
    \begin{figure}
        \centering
        \includegraphics[]{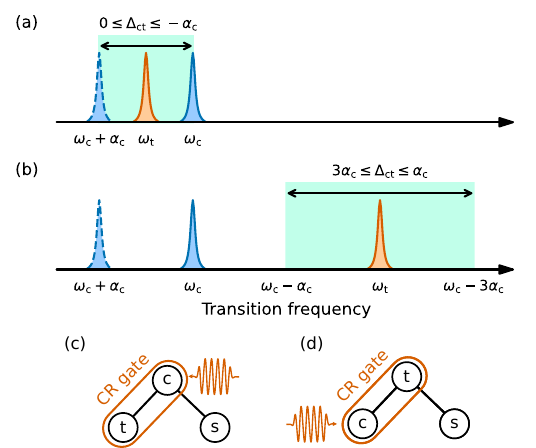}
        \caption{Energy-level diagrams of the control and target qubits used in the CR gate. (a)~Straddling regime, where the target-qubit frequency (orange) lies between the $\ket{0}\!\Leftrightarrow\!\ket{1}$ and $\ket{1}\!\Leftrightarrow\!\ket{2}$ transitions of the control qubit (blue). (b) Far-detuned regime, where the control-qubit frequency lies below the $\ket{1}\!\Leftrightarrow\!\ket{2}$ transition of the target qubit. (c) t-c-s connection topology. (d) c-t-s connection topology.}
        \label{fig:level_diagram}
    \end{figure}

    The CR gates are typically implemented in the so-called straddling regime, where the target-qubit frequency lies between the $\ket{0}\!\Leftrightarrow\!\ket{1}$ and $\ket{1}\!\Leftrightarrow\!\ket{2}$ transitions of the control qubit. This condition enhances the effective $ZX$ interaction while suppressing residual $ZZ$ interactions. However, it tightly restricts the designable control--target detuning $\Delta_{\mathrm{ct}} = \omega_{\mathrm{c}} - \omega_{\mathrm{t}}$ to a narrow interval set by the control-qubit anharmonicity $\alpha_{\mathrm{c}}$ [Fig.~\ref{fig:level_diagram}(a)],
    \begin{align}
        0 \leq \Delta_{\mathrm{ct}} \leq -\alpha_{\mathrm{c}}.
    \end{align}

    As device size grows, the tight parameter conditions required for straddling CR gates lead to severe frequency crowding~\cite{hertzberg2021laser,zhang2022high,malekakhlagh2020first}, making it challenging to scale the current CR gate designs. Indeed, perturbative collision analyses of the straddling CR gate have shown that even with a 1\% gate-error budget, the probability of obtaining a collision-free chip becomes vanishingly small if the system size approaches the thousand-qubit scale~\cite{hertzberg2021laser}. At the same time, surface code architectures require two-qubit gate errors on the order of $10^{-3}$ to implement a logical qubit with reasonable overhead~\cite{fowler2012surface, litinski2019game}. These limitations motivate us to revisit the CR gate in an unexplored parameter regime that relaxes the detuning constraint at the cost of reduced two-qubit gate efficiency, i.e., the ratio between the gate speed and drive amplitude.

    Operating outside the straddling regime does not mean that qubit frequencies can be chosen arbitrarily. Hardware constraints still impose practical limits, since an excessively large frequency difference between qubits would require an impractically large drive amplitude to implement the CR gate within a reasonable gate time. With these constraints in mind, we define the far-detuned regime, illustrated in Fig.~\ref{fig:level_diagram}(b), as the parameter range satisfying
    \begin{align}
        3\alpha_{\mathrm{c}} \leq \Delta_{\mathrm{ct}} \leq \alpha_{\mathrm{c}}.
    \end{align}
    Although $\Delta_{\mathrm{ct}}$ is still bounded, this condition allows a qubit--qubit detuning that is twice as large as in the straddling regime and provides greater flexibility in frequency allocation. The larger detuning also suppresses residual $ZZ$ interactions, albeit not as strongly as in the straddling regime, and can yield gate errors below 0.1\%. Therefore, far-detuned CR gates can reduce the risk of frequency collisions and provide a promising route toward scalable fixed-frequency transmon processors.

    \subsection{$ZX$ coefficient}
        \begin{figure}[t]
            \centering
            \includegraphics[]{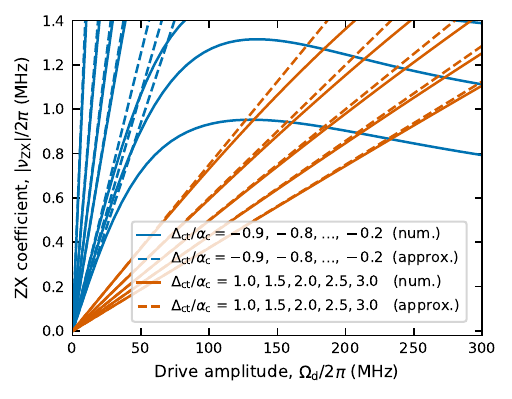}
            \caption{$ZX$ coefficient of the CR gate under a constant-amplitude drive. The $ZX$ coefficient is plotted as a function of the drive amplitude $\Omega_{\mathrm{d}}$ for various detuning ratios $\Delta_{\mathrm{ct}}/\alpha_{\mathrm{c}}$ within the straddling regime ($-1 \leq \Delta_{\mathrm{ct}}/\alpha_{\mathrm{c}} \leq 0$) and far-detuned regime ($1 \leq \Delta_{\mathrm{ct}}/\alpha_{\mathrm{c}} \leq 3$). The solid curves show numerical results using simulation methods in Ref.~\citenum{tripathi2019operation}. The dashed curves show the perturbative prediction in Eq.~\eqref{eq:nuZX}. Parameters used in this simulation are as follows: $\alpha_{\mathrm{c}}/2\pi=\alpha_{\mathrm{t}}/2\pi=-350~\mathrm{MHz}$ and $\abs{g_{\mathrm{ct}}/\Delta_{\mathrm{ct}}}=0.015$.}
            \label{fig:powervsgatetime}
        \end{figure}
        To compare the CR gate operation in the far-detuned regime with that in the straddling regime, we derive the effective $ZX$ interaction rate under a constant drive. A second-order Schrieffer-Wolff transformation yields an effective Hamiltonian that contains a $ZX$ term with coefficient~\cite{tripathi2019operation,magesan2020effective}
        \begin{align}
            \nu_{ZX} = 
            \frac{g_{\mathrm{ct}}\alpha_{\mathrm{c}}\Omega_{\mathrm{d}}}{\Delta_{\mathrm{ct}}\!\left(\Delta_{\mathrm{ct}}+\alpha_{\mathrm{c}}\right)},
            \label{eq:nuZX}
        \end{align}
        where $\Omega_{\mathrm{d}}$ is the drive amplitude on the control qubit, and $g_{\mathrm{ct}}$ is the coupling strength between the control and target qubits.
        
        Numerical simulations of $\nu_{ZX}$ in Fig.~\ref{fig:powervsgatetime} show that the analytic expression remains accurate for the parameter range that we consider in this work, namely $\nu_{ZX}/2\pi \lesssim 1~\mathrm{MHz}$. We note that near $\abs{\Delta_{\mathrm{ct}}/\alpha_{\mathrm{c}}} \approx 0$, the perturbative expression overestimates the attainable $\nu_{ZX}$ but this estimation bias does not adversely affect our comparison between the straddling and far-detuned regimes in the frequency-collision analysis. We therefore use Eq.~\eqref{eq:nuZX} to determine the drive amplitude required for a given gate speed and set of qubit parameters.

    \subsection{Implications of strong-drive operation}
        A larger control--target detuning $\Delta_{\mathrm{ct}}$ allows more flexible frequency allocation, but at the same time reduces the strength of the CR interaction. Consequently, maintaining a sufficiently short gate time in this regime requires a larger drive amplitude. Figure~\ref{fig:powervsgatetime} illustrates this trade-off. To achieve $\nu_{ZX}/2\pi = 1~\mathrm{MHz}$, the numerical simulation predicts a required drive amplitude of $\Omega_{\mathrm{d}}/2\pi \simeq 200~\mathrm{MHz}$ in the far-detuned regime, compared with $\Omega_{\mathrm{d}}/2\pi \simeq 20~\mathrm{MHz}$ in the straddling regime.
        
        Under such strong drives, off-resonant excitation of the control qubit can become non-negligible. As a simple estimate, consider driving the control qubit at the target-qubit frequency with amplitude $\Omega_{\mathrm{d}}$ during a CR gate. Assuming a square pulse waveform, the control qubit undergoes a detuned Rabi oscillation with detuning $\Delta_{\mathrm{ct}}$ and the probability of finding the control qubit in the excited state is given by
        \begin{align}
            P_{\mathrm{e}}(t)
            = \frac{\Omega_{\mathrm{d}}^{2}}{\Delta_{\mathrm{ct}}^{2} + \Omega_{\mathrm{d}}^{2}}
            \sin^{2}\!\qty(\frac{\sqrt{\Delta_{\mathrm{ct}}^{2}+\Omega_{\mathrm{d}}^{2}}}{2}\,t) \leq \frac{\Omega_{\mathrm{d}}^{2}}{\Delta_{\mathrm{ct}}^{2} + \Omega_{\mathrm{d}}^{2}}, \label{eq:pe_max}
        \end{align}
        where the upper bound gives the maximum excitation probability. For the far-detuned parameters in Fig.~\ref{fig:powervsgatetime}, this estimate yields $P_{\mathrm{e}} \lesssim 0.08$, whereas it is only $\lesssim 0.01$ in the straddling regime. Because of the strong drive, the perturbative approach used in previous works becomes unreliable, as it can violate the weak-drive assumptions and induce higher-order multi-photon transitions that were not considered in the perturbative treatment. Therefore, to evaluate frequency collisions in the far-detuned regime, we need to develop a numerical method that predicts the transition amplitude under smoothly ramped strong drives.

\section{Long-time-averaged method} \label{sec:lta_theory}
    In this section, we develop a framework to numerically simulate state-transition amplitudes under smoothly ramped pulses. We consider a network of capacitively coupled, fixed-frequency transmon qubits and model it as coupled Duffing oscillators. Here and throughout the paper, we set $\hbar = 1$. After transforming to a frame rotating at the drive frequency $\omega_{\mathrm{d}}$ and applying the rotating-wave approximation, we write

    \begin{align}
        \hat H_{\mathrm{q}} &= \sum_{i \in \qty{\mathrm{c}, \mathrm{t}, \mathrm{s}}} \qty(\Delta_i \hat a_i^\dagger \hat a_i + \frac{\alpha_i}{2}\hat a_i^\dagger \hat a_i^\dagger \hat a_i \hat a_i), \\
        \hat H_{\mathrm{g}} &= \sum_{i, j \in \qty{\mathrm{c}, \mathrm{t}, \mathrm{s}}} g_{ij} \left(\hat a_i^\dagger \hat a_j + \hat a_i \hat a_j^\dagger\right),
    \end{align}
    where $\hat H_{\mathrm{q}}$ is the Hamiltonian of the individual qubits, $\hat H_{\mathrm{g}}$ is the coupling Hamiltonian, $\omega_i$ is the qubit frequency, $\Delta_i = \omega_i - \omega_{\mathrm{d}}$ is the detuning between the qubit and drive frequencies, $\alpha_i$ is the anharmonicity, and $g_{ij}$ is the coupling strength between the neighboring qubits. When the control qubit is driven at frequency $\omega_{\mathrm{d}}$ and drive amplitude $\Omega_{\mathrm{d}}(t)$, the same rotating-wave approximation reduces the drive Hamiltonian to
    \begin{align}
        \hat H_{\mathrm{d}}(t) = \frac{\Omega_{\mathrm{d}}(t)}{2} \qty(\hat a_\mathrm{c} + \hat a_\mathrm{c}^\dagger).
    \end{align}
\subsection{Qubit dynamics under a square pulse}
    The system dynamics becomes simple when the drive is applied as an ideal square pulse with constant amplitude $\Omega_{\mathrm{d}}$. In a rotating frame defined by the drive frequency, the total Hamiltonian $H_{\mathrm{tot}} = H_{\mathrm{q}} + H_{\mathrm{g}} + H_{\mathrm{d}}$ is time independent under the microwave drive. Writing a three-qubit Fock state as $\ket{\bm n} = \ket{n_{\mathrm{c}}, n_{\mathrm{t}}, n_\mathrm{s}}$, the time evolution from an initial state $\ket{\bm n}$ is then given by
    \begin{align}
        \ket{\psi(t)}
        &= e^{-i t\hat H_{\mathrm{tot}}}\ket{\bm {n}}
        = \sum_{\bm k} e^{-i\bar{E}_{\bm k} t}
            \ket{\bar{\bm k}}\!\bra{\bar{\bm k}}\ket{\bm {n}},
    \end{align}
    where $\ket{\bar{\bm k}}$ is the eigenstate of the driven Hamiltonian $\hat H_{\mathrm{tot}}$ with eigenenergy $\bar{E}_{\bm{k}}$. The probability of finding the state in $\ket{\bm m}$ at time $t$ is
    \begin{align}
        R_{\bm{m}, \bm{n}}(t)
        & = \abs{\bra{\bm m} \ket{\psi(t)}}^{2} \\
        & = \abs{\sum_{\bm k}
            e^{-i \bar{E}_{\bm k}t}
            \bra{\bm m} \ket{\bar{\bm k}}\!
            \bra{\bar{\bm k}} \ket{\bm n}}^{2}.
    \end{align}
    Averaging over $t$ removes the oscillatory off-diagonal contributions from the terms with $\bar{E}_{\bm n}\neq \bar{E}_{\bm m}$, yielding
    \begin{align}
        R_{\bm{m}, \bm{n}}
        &= \lim_{T\to\infty} \frac{1}{T} \int_{0}^{T} R_{\bm{m}, \bm{n}}(t) \dd{t} \\
        &= \sum_{\bm k}
        \Bigl|\braket{\bm m}{\bar{\bm k}}\!
        \braket{\bar{\bm k}}{\bm n}\Bigr|^{2}.
        \label{eq:lta_transition_probability1}
    \end{align}
\subsection{Qubit dynamics under smooth pulse ramps}
    When the control waveform is not constant, we must explicitly include the time-ordering operator to track the system dynamics. Assuming smooth ramp-up and ramp-down envelopes with ramp duration $\tau$ and flattop duration $T_{\mathrm{top}}$, we may decompose the time evolution into three parts:
    \begin{align}
        \hat{U}^{\mathrm{up}} &= \mathcal{T} \qty[\exp(-i \int_{0}^{\tau} \hat{H}_{\mathrm{tot}}(t) \dd{t})], \\
        \hat{U}^{\mathrm{top}} &= \exp(-i \hat{H}_{\mathrm{tot}} T_{\mathrm{top}}), \\
        \hat{U}^{\mathrm{down}} &= \mathcal{T} \qty[\exp(-i \int_{T_{\mathrm{top}} + \tau}^{T_{\mathrm{top}} +2\tau} \hat{H}_{\mathrm{tot}}(t) \dd{t})],
        \label{eq:transposerel}
    \end{align}
    where $\mathcal{T}$ denotes the time-ordering operator. We note that if the ramp-up and ramp-down envelopes are time-symmetric, the ramp-down propagator is given by $\hat{U}^{\mathrm{down}} = [\hat{U}^{\mathrm{up}}]^{\mathrm{T}}$~\cite{mihov2024qubit}.

    Substituting these operators into Eq.~\eqref{eq:lta_transition_probability1}, we obtain
    \begin{align}
        R_{\bm{m}, \bm{n}}(T_{\mathrm{top}})
        &= \abs{\bra{\bm{m}} \hat{U}^{\mathrm{down}} \hat{U}^{\mathrm{top}} \hat{U}^{\mathrm{up}} \ket{\bm{n}}}^{2} \\
        &= \abs{\sum_{\bm{k}} e^{-i \bar{E}_{\bm{k}} T_{\mathrm{top}}} \bra{\psi_{\bm{m},\,\downarrow}} \ket{\bar{\bm{k}}}\!\bra{\bar{\bm{k}}} \ket{\psi_{\bm{n},\,\uparrow}}}^{2}, \label{eq:lta_trans_ramp}
    \end{align}
    where we define the ramp-up state $\ket{\psi_{\bm{n}, \uparrow}}=\hat{U}^{\mathrm{up}} \ket{\bm{n}}$ and the ramp-down state $\ket{\psi_{\bm{m}, \downarrow}}=[\hat{U}^{\mathrm{down}}]^\dagger \ket{\bm{m}}$. We refer to this quantity $R_{\bm m, \bm n}$ as the long-time-averaged (LTA) transition probability. In the adiabatic limit, the ramp-up state coincides with the drive-dressed state $\ket{\bar{\bm{n}}}$ up to a phase.
    
    Analytically, the LTA transition probability under smooth ramp pulses depends on the Fourier transform of the envelope. See the analytical approximation of the LTA transition probability under raised-cosine flattop pulses in Appendix~\ref{sec:transition_prob}.

\section{Frequency-collision analysis} \label{sec:collision}
    In this section, we use the long-time-averaged (LTA) method and perform a systematic parameter sweep to compute frequency-collision landscapes for the CR gate. We first introduce the simulation conditions in Table~\ref{tab:simulation_params} and then describe how LTA transition probabilities are evaluated for realistic pulse shapes and drive amplitudes. Next, we derive the approximate infidelity expressions that depend only on unwanted state transitions and residual $ZZ$ interaction and explain how these expressions are evaluated while sweeping the qubit frequencies over a wide range of control--target--spectator detunings. We show how these sweep results are visualized to identify collision processes and tolerable parameter windows, which will be used as design guidelines in the subsequent sections.

    \subsection{Parameter-sweep setup}
        \subsubsection{Three-qubit-chain topologies} \label{subsec:three_qubit_topology}
            We study two kinds of three-qubit-chain topologies: target--control--spectator (t-c-s) [Fig.~\ref{fig:level_diagram}(c)] and control--target--spectator (c-t-s) [Fig.~\ref{fig:level_diagram}(d)] connection topologies. For both straddling and far-detuned frequency allocation, we extract both kinds of three-qubit chains from a large-scale lattice.

            In the t-c-s topology, the control qubit is coupled to both the target and spectator qubits. Here, the spectator-qubit frequency is designed in the same frequency range as the target qubit. This chain is characterized by the two detunings, the control--target detuning $\Delta_{\mathrm{ct}}$ and control--spectator detuning $\Delta_{\mathrm{cs}}$, which are chosen to satisfy the detuning constraint depending on the operating regime. 
            
            On the other hand, in the c-t-s topology, the target qubit is coupled to both control and spectator qubits. Here, the spectator-qubit frequency is designed in the same frequency range as the control qubit. This chain is characterized by the two detunings, control--target detuning $\Delta_{\mathrm{ct}}$ and spectator--target frequency detuning $\Delta_{\mathrm{st}}$. 

    \subsubsection{Three-qubit-chain parameters} \label{subsec:param_sweep}
        To systematically identify collision conditions, we perform systematic parameter sweeps over three-qubit-chain parameters. We define the three-qubit-chain parameter $\bm \omega$, which characterizes the chain as
        \begin{align}
            \bm{\omega} = \qty(\omega_{\mathrm{c}}, \omega_{\mathrm{t}}, \omega_{\mathrm{s}}, \alpha_{\mathrm{c}}, \alpha_{\mathrm{t}}, \alpha_{\mathrm{s}}).
        \end{align}

        Here, $\omega_{\mathrm{c}}$, $\omega_{\mathrm{t}}$, and $\omega_{\mathrm{s}}$ are the frequencies of the control, target, and spectator qubits, respectively, and $\alpha_{\mathrm{c}}$, $\alpha_{\mathrm{t}}$, and $\alpha_{\mathrm{s}}$ are their anharmonicities. We use the same definition of $\bm{\omega}$ for both the t-c-s and c-t-s topologies. 
        
        During the sweeps, we fix the anharmonicities, the coupling ratio, and the pulse-shape parameters to the values summarized in Table~\ref{tab:simulation_params}. Since frequency collisions are characterized by relative detunings rather than absolute frequencies, we may fix the control-qubit frequency to $\omega_{\mathrm{c}}/2\pi = 5.0\,\mathrm{GHz}$ without loss of generality. For each parameter $\bm \omega$, we fine-tune the drive frequency $\omega_{\mathrm{d}}$ around the bare target-qubit frequency $\omega_{\mathrm{t}}$ and set it to the frequency that minimizes the energy difference between the $\ket{000}$ and $\ket{010}$ states in the rotating frame.

        \begin{table}
            \centering
            \caption{Summary of simulation parameters.}
            \renewcommand{\arraystretch}{1.15}
            \begin{tabular}{llr}
                \hline \hline
                \text{Parameter} & \text{Symbol} & \text{Value} \\
                \hline
                Anharmonicity & $\alpha_{\mathrm{c},\mathrm{t},\mathrm{s}}/2\pi$ & $-350~\mathrm{MHz}$ \\
                Flattop duration & $T_{\mathrm{top}}$ & $250~\mathrm{ns}$ \\
                Ramp duration & $\tau$ & $35~\mathrm{ns}$ \\
                Coupling ratio & $\abs{g_{\mathrm{ct}}/\Delta_{\mathrm{ct}}}$ & $0.015$ \\
                \hline \hline
            \end{tabular}
            \label{tab:simulation_params}
        \end{table}
       
    \subsubsection{Infidelity estimate}
        We estimate the gate error for each three-qubit-chain parameter $\bm \omega$ by combining two contributions: (i) unwanted population transfer $\epsilon_{\mathrm{pop}}$ and (ii) coherent errors due to residual $ZZ$ interaction $\epsilon_{ZZ}$. To quantify errors from unwanted population transfer during the CR gate we use
        \begin{align}
            \epsilon_{\mathrm{pop}} &= \frac{1}{2^n+1} \sum_{(\bm m, \bm n) \in \mathcal{S}^{\perp}_{ZX}(n)} R_{\bm m, \bm n}  \nonumber \\
            &\qquad + \frac{1}{2^n} \sum_{(\bm m, \bm n) \in \mathcal{S}_{\mathrm{leak}}(n)}  R_{\bm m, \bm n}, \label{eq:infidelity}
        \end{align}
        where $n = 3$ is the number of qubits and $R_{\bm m, \bm n}$ denotes the LTA transition probability from the initial state $\ket{\bm n}$ to the final state $\ket{\bm m}$. We denote the set of state transitions that remain in the computational subspace but deviate from the ideal CR interaction as $\mathcal{S}^{\perp}_{ZX}(n)$, and the set of state transitions that leak out of the computational subspace as $\mathcal{S}_{\mathrm{leak}}(n)$. See Appendix~\ref{app:infidelity_derivation} for the detailed definition of these sets and the derivation of this formula.
        

        We also include the error contribution from the residual $ZZ$ interaction. For each qubit parameter set $\bm \omega$, we numerically compute the residual $ZZ$ interaction $\nu_{ZZ}$. Assuming the rotary CR waveform, we obtain the gate-infidelity contribution from the coherent $ZZ$ error $\epsilon_{ZZ}$. See Appendix~\ref{sec:zz_error} for more detailed derivations.
        
        The total infidelity is expressed as the sum of these two contributions,
        \begin{align}
            \epsilon_{\mathrm{total}} = \epsilon_{\mathrm{pop}} + \epsilon_{ZZ},
        \end{align}
        which we use to evaluate the gate error for each three-qubit-chain parameter $\bm \omega$ during the parameter sweep in the following analysis.

        \begin{figure*}
            \centering
            \includegraphics[]{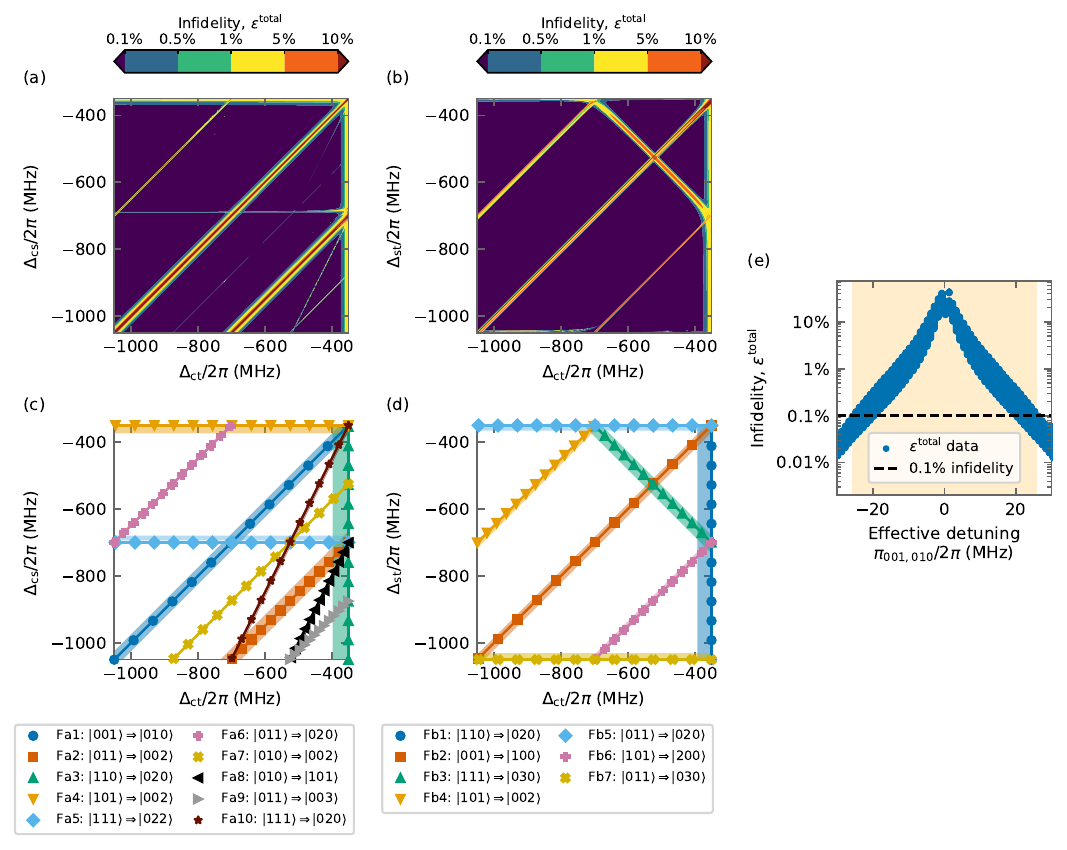}
            \caption{
            Frequency-collision landscapes and frequency-collision maps for three-qubit chains in the far-detuned regime at the $0.1\%$ infidelity threshold.
            (a) and~(b) Frequency-collision landscapes for the target--control--spectator (t-c-s) and control--target--spectator (c-t-s) topologies, respectively. Bright regions indicate where the total infidelity $\epsilon_{\mathrm{total}}$ exceeds $0.1\%$. Qubit frequencies can be assigned in the dark regions that avoid all collisions.
            (c) and~(d) Frequency-collision maps constructed from the collision landscapes in (a) and (b) at the $0.1\%$ infidelity threshold for the t-c-s and c-t-s topologies, respectively. The highlighted regions indicate where the infidelity exceeds $0.1\%$, and the solid center lines indicate the on-resonance conditions $\pi_{\bm m,\bm n}(\bm{\omega})=0$.
            (e)~One-dimensional projection of the collision process $\pi_{001,010}(\bm{\omega})$. Blue points show $\epsilon_{\mathrm{total}}$ as a function of $\pi_{001,010}/2\pi$, and the orange shaded region shows the range of projection values for which $\epsilon_{\mathrm{total}} > 0.1\%$, defining the collision bandwidth for this process.
            }
            \label{fig:collmap_riken}
        \end{figure*}

        \subsubsection{Drive-power optimization via ac Stark shift}
            We perform a simple optimization that mimics the experimental fine-tuning of the CR drive power. The ac Stark shift provides a practical control knob for suppressing narrow frequency collisions during a CR operation. While the ac Stark shift scales approximately quadratically with the drive amplitude $\Omega_{\mathrm{d}}$, the effective $ZX$ term scales only linearly. A small fractional change in drive power can therefore move parasitic transitions off resonance while only weakly perturbing the gate speed. This mechanism is particularly useful in the far-detuned regime, where the stronger drive leads to a larger ac Stark shift for the same fractional change in $\Omega_{\mathrm{d}}$.

            For each three-qubit-chain parameter $\bm{\omega}$, we first compute the analytical drive amplitude from Eq.~\eqref{eq:nuZX} for the gate time $T_{\mathrm{gate}}$. We then sweep the drive amplitude by $\pm 5\%$ around this analytical value and, for each trial amplitude, we evaluate the total infidelity using the procedure described above. The power-optimized infidelity $\epsilon_{\mathrm{total}}$ is used as the total infidelity for the parameter set $\bm{\omega}$. The infidelity landscapes after the parameter sweep for the t-c-s and c-t-s topologies are shown in Figs.~\ref{fig:collmap_riken}(a) and~(b), respectively. Nevertheless, we note that not all collisions can be removed in this way. Static (drive-frequency-independent) collisions as well as strongly driven broad collision processes whose linewidth exceeds the available ac-Stark-shift tuning range cannot be mitigated by this method.

        \subsection{Frequency-collision landscapes}
            The two-dimensional parameter sweeps yield infidelity landscapes for the t-c-s [Fig.~\ref{fig:collmap_riken}(a)] and c-t-s [Fig.~\ref{fig:collmap_riken}(b)] topologies in the far-detuned regime. In both figures, bright regions indicate parameter sets where the total infidelity $\epsilon_{\mathrm{total}}$ exceeds the $0.1\%$ threshold, while the dark background indicates regions below the threshold where qubit frequencies can be allocated with sufficiently low error. From these landscapes, we identify 10 distinct collision processes in the t-c-s topology and 7 in the c-t-s topology. This difference is explained by the connectivity around the driven control qubit. In the t-c-s topology, both the target and spectator are nearest neighbors of the control, which facilitates three-qubit state transitions and leads to more types of collision processes. In contrast, in the c-t-s topology the spectator is a next-nearest-neighbor of the control, suppressing such three-qubit transition processes.

            The infidelity landscapes exhibit several notable features. First, frequency collisions appear as finite-width bands, often approximately straight lines. This structure reflects the fact that each collision is characterized by an effective detuning between the CR drive tone and a collision process, which depends, to leading order, linearly on the device parameters. Second, some collision processes are horizontal or vertical in the two-dimensional landscape. These lines correspond to collision processes whose excitation numbers are conserved after the state transition, which we refer to as static collisions. Static collisions do not depend on the drive frequency or drive amplitude, so they are harder to mitigate by drive amplitude optimization. Our numerical analysis reveals several narrow collision processes that do not appear in previous analyses of straddling-regime CR. These collision processes include next-nearest-neighbor state transitions ($\ket{010} \Rightarrow \ket{002}$ and $\ket{011} \Rightarrow \ket{003}$) and three-body state transitions ($\ket{111} \Rightarrow \ket{020}$ and $\ket{011} \Rightarrow \ket{003}$) in the t-c-s topology. For the c-t-s topology, we observe a three-qubit $ZZZ$ frequency collision $\ket{111} \Rightarrow \ket{030}$ as previously pointed out in Ref.~\citenum{xu2024lattice}. The parameter sweep results within the straddling regime are summarized in Appendix~\ref{app:straddling_collision_map}.

    \subsection{Frequency-collision maps}
        The fact that collision processes exhibit a band-like structure in Figs.~\ref{fig:collmap_riken}(a) and~(b) allows us to associate each collision process with an on-resonance line and the corresponding collision window in the infidelity landscape. By excluding these unusable regions due to collisions, we obtain a parameter space where qubit frequencies can be allocated while keeping the gate infidelity below a chosen threshold. 

        Consider a state transition $\ket{\bm m}\!\Rightarrow\!\ket{\bm n}$ and let $\Delta N$ denote the change in total excitation number. We define the following quantity, which evaluates how close the CR drive is to resonance with the transition, and refer to it as the effective detuning:
        \begin{align}
            \pi_{\bm m,\bm n}(\bm \omega)
            &= (E_{\bm m}-E_{\bm n}) - \Delta N (E_{010}-E_{000}) \nonumber \\
            &= \bm a^{\mathrm{T}}_{\bm m,\bm n} \bm \omega, \label{eq:proj-def-correct}
        \end{align}
        where $E_{\bm n}$ denotes the bare energy of the state $\ket{\bm n} = \ket{n_{\mathrm{c}}, n_{\mathrm{t}}, n_{\mathrm{s}}}$ obtained from the Duffing oscillator model, and $n_{\mathrm{c}}$, $n_{\mathrm{t}}$, and $n_{\mathrm{s}}$ denote the excitation numbers of the control, target, and spectator qubits, respectively. This notation is used for both the t-c-s and c-t-s topologies. In the second line, we expand the expression in terms of the device parameters, where we use the fact that the effective detuning $\pi_{\bm m,\bm n}(\bm \omega)$ is expressed by the product of parameter vector $\bm \omega$ and the coefficient vector 
        \begin{align}
            \bm a_{\bm m,\bm n} = \qty(
                a^{\omega, \mathrm{c}}_{\bm m,\bm n}, a^{\omega, \mathrm{t}}_{\bm m,\bm n}, a^{\omega, \mathrm{s}}_{\bm m,\bm n}, a^{\alpha, \mathrm{c}}_{\bm m,\bm n}, a^{\alpha, \mathrm{t}}_{\bm m,\bm n}, a^{\alpha, \mathrm{s}}_{\bm m,\bm n}
            ),
        \end{align}
        where the first (second) half of the coefficients corresponds to the qubit frequencies (anharmonicities) of the control, target, and spectator qubits, respectively. The coefficient vector $\bm a_{\bm m,\bm n}$ can be uniquely determined from Eq.~\eqref{eq:proj-def-correct} and the bare energy expression of the Duffing oscillator model:
        \begin{align}
            E_{\bm {m}} = \sum_{i  = \mathrm{c}, \mathrm{t}, \mathrm{s}} \qty[\omega_i m_i + \frac{\alpha_i}{2} m_i (m_i - 1)].
        \end{align}

        We note that $\pi_{\bm m,\bm n}(\bm \omega)=0$ corresponds to the center lines in \sii{Figs.~\ref{fig:collmap_riken}(a), (b), (c) and (d)}, indicating the on-resonance condition between the CR drive and the collision process.
        
        To further simplify the set of collision processes, we group together transitions whose collision vectors are proportional. Here, we represent a transition  $\ket{\bm m}\!\Rightarrow\!\ket{\bm n}$ as $(\bm m,\bm n)$ and identify two collision processes $(\bm m,\bm n)$ and $(\bm m',\bm n')$ as equivalent if their coefficient vectors are proportional:
        \begin{equation}        
            (\bm m,\bm n) \sim (\bm m',\bm n')
            \, \Leftrightarrow \,
            \exists\,\lambda \neq 0 \ \text{s.t.}\ \bm a_{\bm m,\bm n} = \lambda \bm a_{\bm m',\bm n'},
        \end{equation}
        so that state transitions related by $(\bm m,\bm n) \sim (\bm m',\bm n')$ can be treated as a single collision process. After this identification, we denote each distinct collision process as $p$ and the set of such processes as $\mathcal{P}$. The collision processes relevant to the far-detuned and straddling regimes are summarized in Appendix~\ref{app:full-threshold-table}.

        For each collision process $p$, we evaluate the effective detuning $\pi_p(\bm\omega)$ from Eq.~\eqref{eq:proj-def-correct} at every grid point of the parameter sweep. Projecting the infidelity data from Figs.~\ref{fig:collmap_riken}(a) and (b) onto this one-dimensional axis typically shows a sharply localized peak near $\pi_p(\bm\omega) \approx 0$ that decays rapidly as $|\pi_p(\bm\omega)|$ increases, as illustrated in \sii{Fig.~\ref{fig:collmap_riken}(e)} for the collision process \sii{$\ket{001}\Rightarrow\ket{010}$}. We therefore define the collision window as the interval in which the infidelity exceeds a chosen threshold $\epsilon$, and denote its lower and upper bounds by $l_p^{(\epsilon)}$ and $u_p^{(\epsilon)}$, respectively.
        
        We treat a three-qubit parameter $\bm\omega$ as having a frequency collision if there exists a process $p \in \mathcal{P}$ such that
        \begin{align}
            \exists p \in \mathcal{P} \ \text{s.t.}\ l_p^{(\epsilon)} \;<\; \pi_p(\bm\omega) \;<\; u_p^{(\epsilon)}, \label{eq:collisioncond}
        \end{align}
        where we note that the bounds $l_p^{(\epsilon)}$ and $u_p^{(\epsilon)}$ are not necessarily symmetric about zero because of the ac Stark shift and qubit--qubit coupling effects.

        In practice, multiple collision processes can overlap for some parameter regions and distort the infidelity peak shapes, which makes it difficult to cleanly determine the collision bounds and can produce stray collision points in the collision landscape. We therefore identify dominant collision processes from the sweep and apply minimal adjustments to the bounds $l_p^{(\epsilon)}$ and $u_p^{(\epsilon)}$ so that any remaining stray collision points are absorbed in the adjusted bounds. After this adjustment, we obtain the frequency collision table, a set of lower and upper bounds that can be used as design guidelines to avoid frequency collisions:
        \begin{align}
            \Lambda^{(\epsilon)} = \left\{ (p,\; l_{p}^{(\epsilon)},\; u_{p}^{(\epsilon)})\; : \; p \in \mathcal{P} \right\}. 
        \end{align}

        We create collision tables for infidelity thresholds of $\epsilon \in \{0.1\%, 0.5\%, 1.0\%\}$. See Appendix~\ref{app:full-threshold-table} for the full list of collision tables. Figures~\ref{fig:collmap_riken}\sii{(c) and (d)} provide a simplified visualization, frequency-collision map, of the extracted collision processes and their collision bands. Each center line corresponds to the on-resonance condition $\pi_{p}(\bm \omega) = 0$ identified in Figs.~\ref{fig:collmap_riken}(a) and (b), while the colored bands indicate the parameter region falling into the collision condition in Eq.~\eqref{eq:collisioncond}.

\section{Yield rate of large-scale devices} \label{sec:yield}
   \begin{figure}
        \centering
        \includegraphics[]{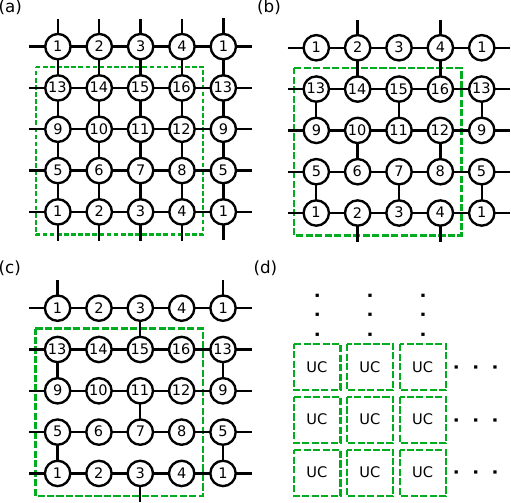}
        \caption{Unit-cell (UC) lattices analyzed in this work. (a)\protect\nobreakdash--(c)~Unit-cell configurations of the square, hexagonal, and heavy-hexagonal lattices, respectively. Each node represents a physical qubit and each edge indicates a capacitive coupling. The node labels identify the frequency group assigned within the unit cell, outlined by the dashed boxes. (d) Scalable lattices are generated by repeating the unit cell.}
        \label{fig:lattices}
    \end{figure}
    In this section, we use the frequency-collision tables derived in Sec.~\ref{sec:collision} to quantify the collision-free yield of large-scale QPUs. We formulate the frequency-allocation optimization as a \sii{mixed-integer} programming problem on a $4\times4$ unit cell with periodic boundary conditions. Using the optimized frequency allocation on the lattice, we simulate the collision-free chip yield of a $1024$-qubit device.

    \subsection{Frequency-allocation optimization}
        We consider periodic $4\times 4$ unit cells of the square, hexagonal, and heavy-hexagonal lattices shown in Fig.~\ref{fig:lattices}. Each node represents a physical qubit, and each edge indicates a capacitive coupling. The optimization variables are the qubit frequencies at each node $\{\omega_i\}$ with one qubit frequency fixed at $\omega_{1}/2\pi = 6000~\mathrm{MHz}$. In addition, we may assign $\omega_{1}$ to a control qubit and all neighboring qubits to target qubits, and extend this assignment throughout the lattice without loss of generality.

        For each lattice topology and operating regime, we impose detuning constraints on every control--target edge and apply periodic boundary conditions by wrapping opposite sides of the unit cell. Using the frequency collision tables $\Lambda^{(\epsilon)}$, we compute the collision margin to the nearest collision band for each three-qubit chain. The optimal frequency allocation is then obtained by solving a \sii{mixed-integer} programming problem that maximizes the smallest margin over all chains and collision processes.

        \subsubsection{Standardized collision margin}
            \begin{figure}
                \centering
                \includegraphics[width=1.0\columnwidth]{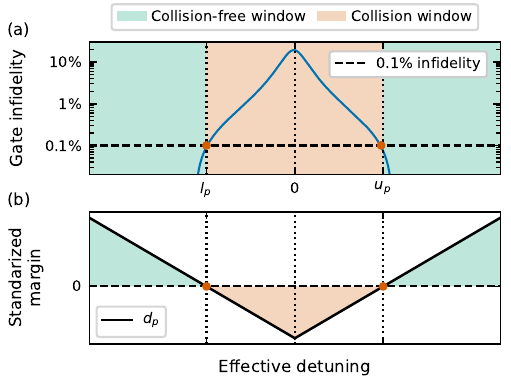}
                \caption{Schematic description of the standardized margin $d_p$ as a function of the effective detuning. (a)~Gate infidelity versus effective detuning. The shaded orange region indicates the collision window where the infidelity exceeds the collision threshold. The lower and upper bounds of this region are defined in Eq.~\eqref{eq:collisioncond}. (b)~Effective detuning as a function of standardized margin $d_p$. The collision window in (a) is mapped to $d_p<0$, while $d_p\ge 0$ corresponds to operation outside the collision condition.}
                \label{fig:dgdefinition}
            \end{figure}
            To model the effect of fabrication-induced qubit-frequency spread on the collision margin, we assume that each qubit frequency deviates from its design value $\omega_i$ by an independent Gaussian shift $\Delta_i$ with standard deviation $\sigma_{\mathrm{f}}$. After the random shifts, we denote the qubit frequency realization as $\qty{\omega_{i}^{*}}$, where $\omega_{i}^{*}$ is the frequency of qubit $i$ after adding a random shift drawn from the Gaussian distribution:
            \begin{align}
                \Delta_i \sim \mathcal{N}(0,\sigma_{\mathrm{f}}^2), \qquad \omega_{i}^{*} = \omega_i + \Delta_i. \label{eq:mc-draw}
            \end{align}
            
            Here, we assume that the qubit anharmonicity, determined primarily by its shunt capacitance, has no fabrication-induced spread. The effective detuning $\pi_p(\bm\omega)$ is a linear combination of qubit frequencies, so its uncertainty is given by the sum of the squared coefficients in the linear combination. If we write $\pi_p(\bm\omega) = \bm a^{\mathrm{T}}_{p} \bm \omega$, then the uncertainty in $\pi_p$ due to qubit-frequency spread is
            \begin{align}
                \delta \pi_{p} = \sigma_{\mathrm{f}}\sqrt{(a^{\omega,\mathrm{c}}_{p})^2 + (a^{\omega,\mathrm{t}}_{p})^2 + (a^{\omega,\mathrm{s}}_{p})^2}. \label{eq:proj-uncertainty}
            \end{align}
            
            It is therefore useful to define a quantity that allows collision margins from different processes to be compared on an equal footing. We define the standardized margin $d_p$ as the normalized distance from the effective detuning $\pi_p(\bm{\omega})$ to the nearest collision bound:
            \begin{align}
                d_{p} (\bm \omega)
                &= \frac{\max \qty[
                    l_{p} - \pi_{p}(\bm \omega),
                    \,
                    \pi_{p}(\bm \omega) - u_{p}
                ]}{\sqrt{(a^{\omega,\mathrm{c}}_{p})^2 + (a^{\omega,\mathrm{t}}_{p})^2 + (a^{\omega,\mathrm{s}}_{p})^2}}, \label{eq:dp_def}
            \end{align}
            where $l_p$ and $u_p$ are the lower and upper bounds of the collision window for process $p$ and are shown in Fig.~\ref{fig:dgdefinition}(a).

            Here, the standardization ensures that the uncertainty is $\delta d_p = \sigma_{\mathrm{f}}$ for all processes, enabling fair comparisons across different collision processes. The collision condition introduced in Eq.~\eqref{eq:collisioncond} is then equivalent to $d_{p}(\bm \omega) < 0$. Conversely, a negative margin indicates that the three-qubit chain $\bm \omega$ has a frequency collision with process $p$ [Fig.~\ref{fig:dgdefinition}~(b)].

        \subsubsection{Objective function}
            We formulate the frequency allocation as a max--min optimization problem, aiming to maximize the smallest normalized margin over all three-qubit chains in a lattice. The optimization problem can be expressed as follows:
            \begin{align}
                \label{eq:maxmin_compact}
                \underset{\qty{\omega_i}}{\text{Maximize }}\quad & z \nonumber \\
                \text{Subject to}\quad & z \le d_p(\bm\omega_{i,j,k}), \hspace{3.9em} \forall p\in\mathcal{P}, (i,j,k)\in \mathcal{E}_3, \nonumber\\
                & \Delta_{\mathrm{ct}}^{\mathrm{(min)}} \le \omega_{\mathrm{c}, i} - \omega_{\mathrm{t}, j} \le \Delta_{\mathrm{ct}}^{\mathrm{(max)}}, \hspace{0.8em} \forall (i,j) \in \mathcal{E}_{2},
            \end{align}
            where $\bm{\omega}_{i,j,k}$ is the three-qubit chain parameter of control qubit $i$, target qubit $j$, and spectator qubit $k$. $\mathcal{P}$ is the set of all distinct collision processes, $\mathcal{E}_{2}$ is the set of connected control--target qubit pairs and $\mathcal{E}_{3}$ is the set of three-qubit chains, and $\Delta_{\mathrm{ct}}^{\mathrm{(min)}, \mathrm{(max)}}$ are the minimum and maximum control--target detuning constraints depending on the operating regime. The optimization variable $z$ is the objective variable that represents the minimum normalized collision margin. The second constraint enforces detuning constraints between neighboring qubit pairs. For the square, hexagonal, and heavy-hexagonal lattices with periodic boundary conditions (Fig.~\ref{fig:lattices}), we solve the \sii{mixed-integer} programming problem using the Gurobi optimizer~\cite{gurobiopt}. \si{The optimized frequency allocations for the far-detuned and straddling regimes are shown in Figs.~\ref{fig:optfreq_fardetuned} and \ref{fig:optfreq_straddling}, respectively.}

    \subsection{Qubit-frequency spread for 10\% yield}
        We estimate the collision-free yield of a chip with an optimized frequency allocation by directly sampling chip realizations using Monte Carlo simulation. For each Monte Carlo sample, we sample a qubit frequency realization $\qty{\omega_{i}^*}$ by independently adding a random frequency shift $\Delta_{i}$ drawn from a Gaussian distribution in Eq.~\eqref{eq:mc-draw}. For each sample, we apply the same frequency-collision checks for every three-qubit chain on the chip. If any chain yields a margin $d_{p} (\bm \omega) < 0$, the sample is recorded as a collision. The yield is then estimated as the fraction of collision-free samples over $5\times 10^5$ independent trials.
        
    \subsubsection{Required qubit-frequency spread for a 1024-qubit device}
        \begin{table}
        \sii{
            \centering
            \caption{Qubit-frequency spread requirements for achieving a 10\% collision-free yield in a 1024-qubit QPU. Here, $\sigma_{10\%}$ denotes the standard deviation of the qubit-frequency distribution required to obtain 10\% collision-free chips, and $z^{\mathrm{opt}}$ denotes the optimized objective value in Eq.~\eqref{eq:maxmin_compact}.}

            \label{tab:sigma10}
            \begin{tabular}{llrrrr}
                \hline \hline
                \\[-9pt]
                Regime & Lattice & Infidelity
                & \multicolumn{1}{c}{$\displaystyle \frac{\sigma_{10\%}}{2\pi}$}
                & \multicolumn{1}{c}{$\displaystyle \frac{z^{\mathrm{opt}}}{2\pi}$}
                & \multicolumn{1}{c}{$\displaystyle \frac{\sigma_{10\%}}{z^{\mathrm{opt}}}$} \\[6pt]
                & & &
                \multicolumn{1}{c}{(MHz)} &
                \multicolumn{1}{c}{(MHz)} & \\[2pt]
                \hline
                Straddling & Square & 0.1\% & 0.02 & 0.05 & 0.35 \\
                Straddling & Square & 0.5\% & 0.96 & 2.87 & 0.33 \\
                Straddling & Square & 1.0\% & 1.70 & 5.09 & 0.33 \\
                Straddling & Hexagonal & 0.1\% & 1.39 & 4.01 & 0.35 \\
                Straddling & Hexagonal & 0.5\% & 2.51 & 7.29 & 0.34 \\
                Straddling & Hexagonal & 1.0\% & 3.45 & 10.01 & 0.35 \\
                Straddling & Heavy-hexagonal & 0.1\% & 2.35 & 6.80 & 0.35 \\
                Straddling & Heavy-hexagonal & 0.5\% & 3.22 & 9.51 & 0.34 \\
                Straddling & Heavy-hexagonal & 1.0\% & 4.70 & 13.57 & 0.35 \\
                Far-detuned & Square & 0.1\% & 6.86 & 19.61 & 0.35 \\
                Far-detuned & Square & 0.5\% & 8.60 & 25.10 & 0.34 \\
                Far-detuned & Square & 1.0\% & 9.12 & 26.05 & 0.35 \\
                Far-detuned & Hexagonal & 0.1\% & 8.71 & 25.63 & 0.34 \\
                Far-detuned & Hexagonal & 0.5\% & 9.69 & 26.87 & 0.36 \\
                Far-detuned & Hexagonal & 1.0\% & 14.20 & 42.14 & 0.34 \\
                Far-detuned & Heavy-hexagonal & 0.1\% & 12.92 & 37.48 & 0.35 \\
                Far-detuned & Heavy-hexagonal & 0.5\% & 13.41 & 38.89 & 0.35 \\
                Far-detuned & Heavy-hexagonal & 1.0\% & 17.39 & 50.91 & 0.34 \\
                \hline \hline
            \end{tabular}
            }
        \end{table}
        \begin{figure}
            \centering
            \includegraphics[]{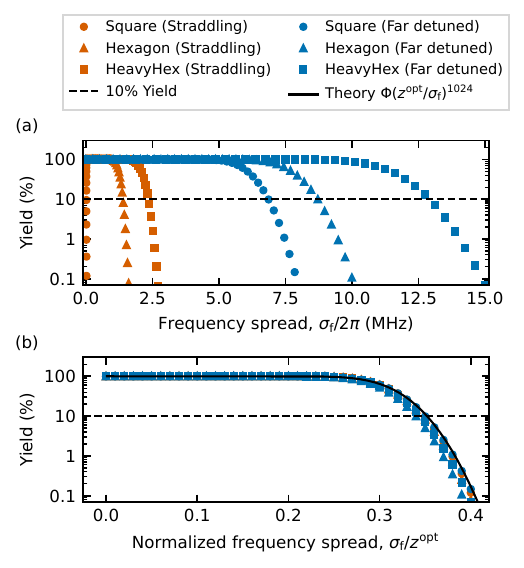}
            \caption{(a) Yield rate of a 1024-qubit chip estimated with a Monte Carlo method as a function of qubit-frequency spread $\sigma_{\mathrm{f}}$. Orange markers indicate the chip yield designed in the straddling regime, while blue markers indicate the chip yield designed in the far-detuned regime. Markers indicate the square, hexagonal, and heavy-hexagonal lattices. (b)~Same yield curves scaled by their optimized objective values $z^{\mathrm{opt}}$ from Table~\ref{tab:sigma10}. All curves show a similar shape when plotted against the normalized qubit-frequency spread $\sigma_{\mathrm{f}}/z^{\mathrm{opt}}$, which can be explained by the independent failure model (see Appendix~\ref{app:yield_derivation}). The solid black line is the theoretical prediction with $N_{\mathrm{active}} = 1024$ [Eq.~\eqref{eq:universal_yield_1024}].}
            \label{fig:yield-01pct}
        \end{figure}
        Fault-tolerant surface-code operations with sufficiently low logical error rates are expected to require a physical gate error rate on the order of $10^{-3}$ and the number of qubits on the order of $10^3$~\cite{fowler2012surface}. To assess how much reduction in qubit-frequency spread is required to obtain a large-scale QPU at 10\% collision-free yield, we estimate the yield of a 1024-qubit device modeled as $8\times8$ repetitions of the optimized $4\times4$ unit cell. We perform a Monte Carlo yield estimate across lattice topologies and tolerable infidelity thresholds. The required qubit-frequency spread values $\sigma_{10\%}$ are summarized in Table~\ref{tab:sigma10}.

        The table shows that achieving a 10\% yield in a 1024-qubit QPU designed in the far-detuned regime requires qubit-frequency spread values of \sii{$\sigma_{10\%}/2\pi=6.86, 8.71, 12.92~\mathrm{MHz}$} for square, hexagonal, and heavy-hexagonal lattices, respectively. Compared with the state-of-the-art value $\sigma_{\mathrm{f}}/2\pi \approx 14.5~\mathrm{MHz}$~\cite{hertzberg2021laser}, this corresponds to an almost twofold reduction in qubit-frequency spread. The straddling design is more demanding, requiring \sii{$\sigma_{10\%}/2\pi=0.02, 1.39, 2.35\,\mathrm{MHz}$} at the same error rate.

        Figure~\ref{fig:yield-01pct}(a) compares the collision-free yield of the straddling and far-detuned designs as a function of qubit-frequency spread. Across all topologies, the far-detuned design exhibits significantly greater tolerance to qubit-frequency spread, resulting in nearly an order of magnitude higher tolerance at the 10\% yield condition. If we further relax the required error rate to 1\%, the far-detuned design on the heavy-hexagonal lattice becomes feasible with the current qubit-frequency spread, with the far-detuned design on the hexagonal lattice near the boundary.
        
        Finally, we estimate the feasible chip sizes achievable with the current state-of-the-art qubit-frequency spread at a 10\% collision-free yield by mapping the qubit-frequency-spread requirement versus chip size and lattice topology at a 0.1\% infidelity requirement. With this current qubit-frequency spread, our simulations predict that far-detuned CR supports up to 32 qubits for the square lattice, \sii{48} qubits for the hexagonal lattice, and 400 qubits for the heavy-hexagonal lattice. See Fig.~\ref{fig:yield_heatmap} in Appendix~\ref{app:yield_derivation} for more details.

    \subsubsection{Yield curve}
        \label{sec:universal_yield}

        The yield curves in Fig.~\ref{fig:yield-01pct}(b) show that the simulated Monte Carlo results collapse approximately onto a common curve when plotted against the normalized qubit-frequency spread $\sigma_\mathrm{f}/z^{\mathrm{opt}}$. This behavior can be understood using a simple independent failure model.

        Our optimization maximizes the minimum standardized distance to the set of collision constraints, as introduced in Sec.~\ref{sec:yield}. The optimal objective value $z^{\mathrm{opt}}$ therefore equals the smallest standardized margin in the design. Equivalently, among the many collision processes included in the constraint set, $z^{\mathrm{opt}}$ represents the standardized distance to the worst-case process.

        We model fabrication-induced qubit-frequency spread as independent Gaussian shifts of each qubit frequency with standard deviation $\sigma_\mathrm{f}$. Because each collision condition is evaluated through a linear projection of qubit frequencies and each constraint is normalized by its projection sensitivity, the standardized margin associated with a given collision process acquires an approximately Gaussian variation with standard deviation $\sigma_\mathrm{f}$. If we assume that the correlation between standardized margins is negligible, then the probability that a worst-case standardized margin remains collision-free (i.e., positive) is given by
        \begin{align}
            p(\sigma_\mathrm{f})
            =
            \frac12\left[1+\mathrm{erf}\!\left(\frac{z^{\mathrm{opt}}}{\sqrt{2} \, \sigma_\mathrm{f}}\right)\right]
            = \Phi\!\left(\frac{z^{\mathrm{opt}}}{\sigma_\mathrm{f}}\right),
            \label{eq:single_qubit_survival}
        \end{align}
        where $\Phi(\cdot)$ denotes the cumulative distribution function of the standard normal distribution.

        Here, if we further assume that chip failure is dominated by such worst-case standardized margins, which we refer to as active margins, the collision-free yield can be approximated as
        \begin{align}
            \mathrm{Yield}(\sigma_\mathrm{f})
            =
            p(\sigma_\mathrm{f})^{N_{\mathrm{active}}}
            =
            \Phi\!\left(\frac{z^{\mathrm{opt}}}{\sigma_\mathrm{f}}\right)^{N_{\mathrm{active}}},
            \label{eq:universal_yield_1024}
        \end{align}
        where $N_{\mathrm{active}}$ is the number of active constraints.

        For the devices studied in this work, we obtain \sii{$N_{\mathrm{active}} = 1096\text{--}2106$}, realized by an $8\times 8$ tiling of the optimized $4\times 4$ unit cell; therefore Eq.~\eqref{eq:universal_yield_1024} gives an analytical expression for the yield rate determined by the dimensionless ratio $\sigma_\mathrm{f}/z^{\mathrm{opt}}$ and the number of active constraints $N_{\mathrm{active}}$.

\section{Discussion and conclusion} \label{sec:discussion}
    \si{In this paper, we developed a systematic framework for frequency-allocation optimization and applied it to cross-resonance (CR) gates in the far-detuned regime. We performed systematic parameter sweeps of the t-c-s and c-t-s three-qubit chains using long-time-averaged (LTA) simulations and identified collision-free parameter conditions at a $0.1\%$ error budget. We then used these conditions as constraints in a mixed-integer programming optimization on square, hexagonal, and heavy-hexagonal lattices and derived optimal frequency allocations that maximize the device yield. Finally, through the Monte Carlo simulations over qubit-frequency spread, we showed that far-detuned CR layouts can tolerate a qubit-frequency spread of $6.86~\mathrm{MHz}$ for the 1024-qubit square lattice at the $0.1\%$ infidelity threshold. Beyond the CR gate, our systematic frequency-allocation optimization framework could also be useful for other two-qubit gates based on fixed-frequency qubits.}

    At the same time, our work has several limitations that suggest interesting directions for future study. First, we have restricted attention to single-tone CR drives, whereas multi-color driving becomes important for parallelized operations such as syndrome extraction in quantum error correction~\cite{heya2024floquet}. Combining the present LTA framework with a Floquet-Hamiltonian analysis~\cite{di2022extensible, heya2024floquet} could enable evaluation of frequency collisions under multi-color drives incorporating the effects of pulse ramping and thus encompass a wider class of experimentally relevant control sequences. Second, our \si{systematic parameter-sweep approach could be extended to other two-qubit gates or qubit--resonator collisions. Possible applications include measurement-induced state transitions~\cite{khezri2023measurement, dumas2024measurement, dai2026characterization} and two-qubit gates with parametric interaction~\cite{nguyen2022blueprint, shirai2023all,shirai2025high}, where detailed frequency-collision studies are now emerging~\cite{ma2025analysis, ai2025scalable}.} We expect that applying the same methodology to these platforms will yield quantitative collision conditions that can be used to create parameter-design guidelines across different gate schemes.

\section*{Acknowledgments}
    We thank A. Noguchi, Z. Wang, S. Wang, T. Miyamura, H.~Kikuchi, and K. Kodama for fruitful discussions. We also acknowledge the RIKEN R-CCS for providing computing resources. S.I. acknowledges the support from the RIKEN Junior Research Associate~(JRA) program and the Forefront Physics and Mathematics Program to Drive Transformation (FoPM), a World-leading Innovative Graduate Study (WINGS) Program, the University of Tokyo. This research was partly supported by the Ministry of Education, Culture, Sports, Science and Technology~(MEXT) Quantum Leap Flagship Program~(Q-LEAP) (Grant No.~JPMXS0118068682) and the Japan Science and Technology Agency (JST) as part of Adopting Sustainable Partnerships for Innovative Research Ecosystem (ASPIRE) (Grant No.~JPMJAP2513). This research was partly funded by JSPS KAKENHI (Grant No.~JP24K23038).

\section*{Data availability}
    The data that support the findings of this study are available from the corresponding author upon reasonable request.

\appendix 

\section{$ZZ$ error estimate} \label{sec:zz_error}
    \begin{figure}
        \centering
        \includegraphics[]{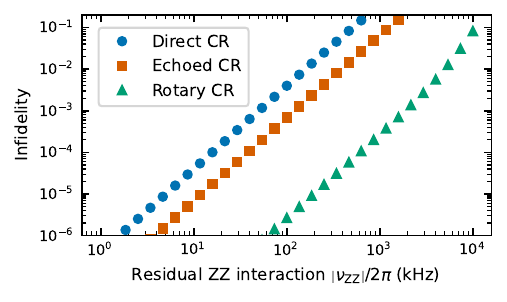}
        \caption{Estimated gate infidelity due to the residual $ZZ$ interaction for three CR waveforms: direct CR, echoed CR, and rotary CR. The infidelity is computed in the computational subspace using an effective two-qubit Hamiltonian with a fixed gate speed $\nu_{ZX}/2\pi=1~\mathrm{MHz}$ ($T_{\mathrm{gate}}=250~\mathrm{ns}$) while sweeping the residual coupling $\abs{\nu_{ZZ}}$.}
        \label{fig:zz_error}
    \end{figure}
    In this section, we compare three cross-resonance (CR) waveforms to quantify the impact of residual $ZZ$ interaction on gate fidelity. The simplest implementation is the direct CR waveform, where a microwave drive is applied to the control qubit at the target-qubit frequency. As a simple mitigation, we consider an echoed CR waveform that inserts an $X_{\pi}$ pulse in the middle of the sequence and reverses the CR drive sign, partially mitigating $ZZ$-induced errors~\cite{corcoles2013process, sheldon2016procedure, takita2017experimental}. Finally, we consider a more sophisticated implementation that applies an additional rotary drive on the target qubit simultaneously with the CR drive on the control qubit, providing stronger suppression of $ZZ$-induced errors; we refer to this as the rotary waveform~\cite{jurcevic2021demonstration, sundaresan2020reducing}.
    We estimate the $ZZ$-induced gate infidelity for the three waveforms as a function of the residual $ZZ$ strength. For simplicity, we analyze infidelities in an effective two-level system Hamiltonian with a fixed entangling $ZX$ rate and a fixed residual $ZZ$ coupling. We employ the parameters summarized in Table~\ref{tab:simulation_params} and set the effective gate speed to $\nu_{ZX}/2\pi = 1~\mathrm{MHz}$. The corresponding gate time is $T_{\mathrm{gate}} = 250~\mathrm{ns}$.

    For the direct CR waveform with residual $ZZ$ interaction $\nu_{ZZ}$, the effective Hamiltonian in the rotating frame is given by
    \begin{align}
        \hat{H}_{\mathrm{eff}} = \nu_{ZX} \frac{\hat{Z}\hat{X}}{2} + \nu_{ZZ} \frac{\hat{Z}\hat{Z}}{2}, & \quad (0 \le t \le T_{\mathrm{gate}}).
    \end{align}

    For an echoed CR waveform, we model the dynamics with the following effective Hamiltonian. In the middle of the sequence, we apply an ideal, instantaneous $\mathrm{X}_{\pi}$ pulse on the control qubit and then flip the sign of the CR drive. The Hamiltonian is given by
    \begin{equation}
        H_{\mathrm{eff}}(t) = 
        \left\{
            \begin{aligned}
                &\nu_{ZX} \frac{\hat{Z}\hat{X}}{2} + \nu_{ZZ} \frac{\hat{Z}\hat{Z}}{2}, \quad (0 \le t < T_{\mathrm{gate}}/2), \\
                &\nu_{ZX} \frac{\hat{Z}\hat{X}}{2} - \nu_{ZZ} \frac{\hat{Z}\hat{Z}}{2}, \quad (T_{\mathrm{gate}}/2 \le t \le T_{\mathrm{gate}}),
            \end{aligned}
        \right.
    \end{equation}
    where in the second half of the Hamiltonian we have used the relation $\hat{X}_{\mathrm{c}} \hat{Z}_{\mathrm{c}}^\dagger \hat{X}_{\mathrm{c}} = -\hat{Z}_{\mathrm{c}}$ to model the effect of the $\mathrm{X}_{\pi}$ pulse on the control qubit. 

    For the rotary CR waveform, we additionally include an $\hat{I}\hat{X}$ term arising from the rotary drive on the target qubit. The effective Hamiltonian is given by
    \begin{equation}
        H_{\mathrm{eff}}(t) =
        \left\{
            \begin{aligned}
                &\nu_{ZX} \frac{\hat{Z}\hat{X}}{2} + \nu_{ZZ} \frac{\hat{Z}\hat{Z}}{2} + \nu_{IX} \frac{\hat{I}\hat{X}}{2}, \\
                &\qquad \qquad \qquad \quad  (0 \le t < T_{\mathrm{gate}}/2),  \\
                &\nu_{ZX} \frac{\hat{Z}\hat{X}}{2} + \nu_{ZZ} \frac{\hat{Z}\hat{Z}}{2} - \nu_{IX} \frac{\hat{I}\hat{X}}{2}, \\
                &\qquad \qquad \qquad \quad (T_{\mathrm{gate}}/2 \le t \le T_{\mathrm{gate}}),
            \end{aligned}
        \right.
    \end{equation}
    where we optimize $\nu_{IX}$ for each residual $ZZ$ interaction strength  $\nu_{ZZ}$ to minimize infidelity.
    
    The gate fidelities of each waveform are computed by simulating the unitary propagator under each Hamiltonian and comparing it with the ideal $ZX_{\pi/2}$ unitary operator. The resulting $ZZ$-induced errors as a function of $\nu_{ZZ}$ are shown in Fig.~\ref{fig:zz_error}. In the main text, we employ the infidelity data obtained using the rotary CR waveform and use cubic-spline interpolation to map a given residual $ZZ$ interaction to its corresponding contribution to the gate infidelity.

\section{Analytical expression for the transition probability} \label{sec:transition_prob}
    We derive an analytical expression for the state-transition probability induced by near-resonant smooth microwave pulses. The analysis follows the framework introduced in~\cite{martinis2014fast, motzoi2013improving, ding2025pulse}, in which the transition amplitude is effectively given by the Fourier transform of the pulse envelope. While the approximation is not valid at resonance, it quickly becomes accurate as the detuning increases.

    Assuming dynamics within a two-level system of interest, we construct a simple Hamiltonian to estimate the transition probability between the two states under microwave drive. Within the subspace spanned by $\{\ket{\psi_\mathrm{a}}, \ket{\psi_\mathrm{b}}\}$, the Hamiltonian is expressed as
    \begin{align}
        \hat H_{\mathrm{eff}}(t) = -\frac{\Delta_{\mathrm{eff}}}{2} \hat{\sigma}_z + \frac{\Omega_{\mathrm{eff}}(t)}{2} \hat{\sigma}_x, \label{eq:two_level_hamiltonian}
    \end{align}
    where $\Delta_{\mathrm{eff}}$ denotes the effective detuning between the two levels under the rotating-wave approximation, and $\Omega_{\mathrm{eff}}(t)$ denotes the effective drive amplitude between the two states. We assume that the drive envelope is smoothly turned on and off, such that
    \begin{align}
        \Omega_{\mathrm{eff}}(0)=\Omega_{\mathrm{eff}}(T_{\mathrm{pulse}})=0,
    \end{align}
    where $T_{\mathrm{pulse}}$ is the total pulse duration including the rise and fall times.

    \begin{figure}
        \centering
        \includegraphics[]{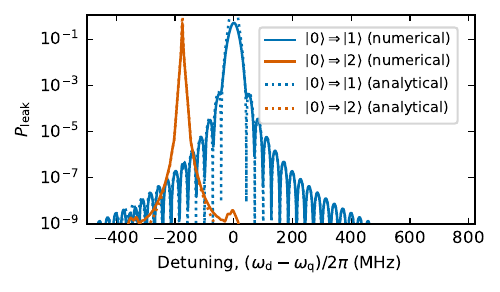}
        \caption{Comparison of the analytical and numerical results for leakage probability under a raised-cosine flattop waveform averaged over the flattop duration $T_{\mathrm{top}}$. The analytical results for the one-photon and two-photon transitions are given by Eqs.~\eqref{eq:leakage_rc_1ph} and~\eqref{eq:leakage_rc_2ph}, respectively, and the numerical results are obtained by evaluating the LTA transition probability in Eq.~\eqref{eq:lta_trans_ramp}. The simulation parameters are $\alpha/2\pi=-350~\mathrm{MHz}$, $\Omega_{0}/2\pi=24~\mathrm{MHz}$, and $\tau=35~\mathrm{ns}$.}
        \label{fig:ltacomparison}
    \end{figure}
    When the effective two-level system is driven with an effective drive amplitude $\Omega_{\mathrm{eff}}(t)$, the instantaneous eigenstates of the time-dependent Hamiltonian are
    \begin{align}
        \ket{\xi_{-}(t)}
        &=\cos\qty(\frac{\varphi(t)}{2})\ket{\psi_\mathrm{a}}
        -\sin\qty(\frac{\varphi(t)}{2})\ket{\psi_\mathrm{b}}, \\
        \ket{\xi_{+}(t)} &= \sin\qty(\frac{\varphi(t)}{2})\ket{\psi_\mathrm{a}} + \cos\qty(\frac{\varphi(t)}{2})\ket{\psi_\mathrm{b}},
    \end{align}
    where we define $\tan \varphi(t)=\Omega_{\mathrm{eff}}(t)/\Delta_{\mathrm{eff}}$.
    
    We consider a quantum state $\ket{\psi(t)}$ evolving under the Hamiltonian in Eq.~\eqref{eq:two_level_hamiltonian} from the initial condition $\ket{\psi(0)}=\ket{\psi_{\mathrm{a}}}$. Because $\Omega_{\mathrm{eff}}(0)=0$, the initial state coincides with the instantaneous eigenstate $\ket{\xi_{-}(0)}$. The state can therefore be expressed using the instantaneous eigenstates as
    \begin{align}
        \ket{\psi(t)} &= \cos\qty(\frac{\abs{\chi(t)}}{2})\ket{\xi_{-}(t)} \nonumber \\
        &\quad + \sin\qty(\frac{\abs{\chi(t)}}{2})e^{i\arg\chi(t)}\ket{\xi_{+}(t)},
    \end{align}
    where $\chi(t)$ is a complex parameter that captures the non-adiabatic mixing between the instantaneous eigenstates. Intuitively, if the system evolution is sufficiently adiabatic, $\ket{\psi(t)}$ follows the instantaneous eigenstate $\ket{\xi_{-}(t)}$ up to an overall phase.

    Assuming that the system evolves nearly adiabatically $\abs{\chi(t)} \ll 1$, we may approximate $\chi(t)$ as
    \begin{align}
        \chi(t) \simeq - \frac{1}{\Delta_{\mathrm{eff}}} \int_{0}^{t} \frac{\dd{\Omega_{\mathrm{eff}}(s)}}{\dd{s}} e^{-i\Delta_{\mathrm{eff}} s} \dd{s}. \label{eq:chi_third}
    \end{align}

    The probability to find the state in $\ket{\psi_{\mathrm{b}}}$ at the end of the pulse at $T_{\mathrm{pulse}}$ is given by
    \begin{align}
        P_{\mathrm{leak}}
        &= \sin^2\!\left(\frac{\abs{\chi(T_{\mathrm{pulse}})}}{2}\right) \\
        &\simeq \frac{1}{4 \Delta_{\mathrm{eff}}^2} \abs{\int_{0}^{T_{\mathrm{pulse}}}  \frac{\dd{\Omega_{\mathrm{eff}}(s)}}{\dd{s}} e^{-i\Delta_{\mathrm{eff}} s} \dd{s}}^{2}. \label{eq:leakage_formula}
    \end{align}
    Although the approximations used to derive this expression assume near-adiabatic conditions, our numerical comparison in Fig.~\ref{fig:ltacomparison} verifies that this approximate expression is a good estimate of leakage probability even around near-resonant conditions.
    
    We present two examples illustrating how a smooth raised-cosine flattop waveform suppresses the broad excitation profile induced by a square pulse. We consider a transmon qubit driven by a microwave pulse with raised-cosine flattop envelope. The system Hamiltonian is given by
    \begin{align}
        H(t) = \omega_{\mathrm{q}}\had \ha + \frac{\alpha}{2} \had \had \ha \ha + \Omega_{\mathrm{d}}(t) \cos(\omega_{\mathrm{d}} t) (\had + \ha),
    \end{align}
    where $\omega_{\mathrm{q}}$ is the qubit frequency, $\alpha$ is the anharmonicity, and $\Omega_{\mathrm{d}}(t)$ is the drive amplitude whose envelope is defined with ramp time $\tau$ and flattop duration $T_{\mathrm{top}}$ as
    \begin{align}
        &\Omega_{\mathrm d}(t) = \Omega_{0} \nonumber \\
        &\times \begin{cases}
        \sin^2\qty(\frac{\pi t}{2\tau}), &(0\le t\le\tau),\\
        1, &(\tau<t<\tau+T_{\mathrm{top}}),\\
        \cos^2\qty[\frac{\pi(t -\tau-T_{\mathrm{top}})}{2\tau}], &(\tau+T_{\mathrm{top}}\le t\le2\tau+T_{\mathrm{top}}),
        \end{cases}
    \end{align}
    where $\Omega_{0}$ is the maximum drive amplitude and the total pulse duration is $T_{\mathrm{pulse}} = 2\tau + T_{\mathrm{top}}$.

    \subsubsection{Leakage of single-photon transitions} \label{app:rc_onephoton}
        We consider the dynamics of a transmon driven at a frequency near its $\ket{0}\!\Leftrightarrow\!\ket{1}$ transition. Assuming that no other states are populated, the dynamics is well described within the $\{\ket{0},\ket{1}\}$ subspace. In the rotating frame, the effective detuning and drive amplitude are
        \begin{align}
            \Delta_{\mathrm{eff}} &= \omega_{\mathrm{d}} - \omega_{\mathrm{q}} = \Delta, \\
            \Omega_{\mathrm{eff}} &= \Omega_{\mathrm{d}}.
        \end{align}

        Substituting these into Eq.~\eqref{eq:leakage_formula}, the time integral can be evaluated analytically. Averaging the final leakage probability over the flattop duration $T_{\mathrm{top}}$, while keeping the ramp time $\tau$ fixed, yields
        \begin{align}
            P_{\mathrm{leak}}
            =\frac{\pi^4 \Omega_{0}^{2}}{2\Delta^{2}} \qty[\frac{\cos(\tfrac{\Delta\tau}{2})}{(\Delta\tau)^{2}-\pi^{2}}]^2. \label{eq:leakage_rc_1ph}
        \end{align}

        Under the raised-cosine waveform, the resulting bound decays as $1/\Delta^{6}$, which means that the off-resonant excitation tails are strongly suppressed by the smooth ramp. This decay is contrasted with the square-pulse excitation profile in Eq.~\eqref{eq:pe_max}, which scales as $1/\Delta^{2}$. Therefore, smooth ramping significantly mitigates the impact of near-resonant frequency collisions.

    \subsubsection{Leakage of two-photon transitions}
        We consider two-photon excitation when a transmon is driven near half of the $\ket{0}\!\Leftrightarrow\!\ket{2}$ transition frequency. Assuming that other states are not populated, we describe the dynamics in the $\{\ket{0},\ket{2}\}$ subspace by adiabatically eliminating the intermediate state $\ket{1}$. The effective detuning and drive amplitude in the rotating frame at $\omega_{\mathrm{d}}$ are
        \begin{align}
            \Delta_{\mathrm{eff}} &= 2 \qty(\omega_{\mathrm{d}} - \omega_{\mathrm{q}} - \frac{\alpha}{2}) = 2\Delta, \\
            \Omega_{\mathrm{eff}} &= \frac{\sqrt{2}\,\Omega_{\mathrm{d}}^{2}}{\alpha},
        \end{align} 
        where we assume that $\Omega_{\mathrm{eff}}$ is independent of the drive frequency.

        Substituting these into Eq.~\eqref{eq:leakage_formula}, we similarly obtain the state-transition probability by averaging the final leakage probability over the flattop duration $T_{\mathrm{top}}$:
        \begin{widetext}
            \begin{align}
                P_{\mathrm{leak}}
                &= \frac{\pi^4\Omega_{\mathrm{d}}^4}{256 \alpha^2 \Delta^2} \times \qty{\frac{\qty[34\pi^4 - 80\pi^2 (\Delta \tau)^2 + 64(\Delta \tau)^4] + 6\pi^2 \qty[5\pi^2 - 8(\Delta \tau)^2] \cos(2 \Delta \tau)}{\qty[\pi^2 - (\Delta \tau)^2]^2 \cdot \qty[\pi^2 - 4(\Delta \tau)^2]^2}}. \label{eq:leakage_rc_2ph}
            \end{align}
        \end{widetext}

        At large detuning, the effective drive amplitude $\Omega_{\mathrm{eff}}$ acquires a non-trivial frequency dependence not captured by the analytical formula, causing the deviation from the numerical results at $P_{\mathrm{leak}} \lesssim 10^{-8}$ in Fig.~\ref{fig:ltacomparison}.

\section{LTA-based infidelity bound} \label{app:infidelity_derivation}
    We derive a population-transfer infidelity metric for a CR gate acting on the three-qubit computational subspace ($n=3$), consisting of the control, target, and spectator qubits.
    
    \subsection{Notation and definitions}
        We denote the set of full three-qubit Fock states as $\mathcal{F}(n)$ and define the computational and non-computational state sets as
        \begin{align}
            \mathcal{F}(n) &= \{\ket{0}, \ket{1}, \ldots \}^{\otimes n}, \\
            \mathcal{C}(n) &= \{\ket{0},\ket{1}\}^{\otimes n}, \\
            \overline{\mathcal{C}}(n) &= \mathcal{F}(n)\setminus \mathcal{C}(n),
        \end{align}
        where we assume that each transmon qubit is truncated at a sufficiently high level such that the truncation does not affect the results. In this work, we truncate each qubit to the four lowest levels, so $\mathcal{F}(n) = \{\ket{0}, \ket{1}, \ket{2}, \ket{3}\}^{\otimes n}$.

        The ideal CR gate implements a $ZX$ rotation of angle $\pi/2$ within the computational subspace. The ideal unitary operator for this operation is
        \begin{align}
            \hat{U}_{ZX} = \frac{1}{\sqrt{2}}
            \begin{pmatrix}
                1 & -i & 0 & 0 \\
                -i & 1 & 0 & 0 \\
                0 & 0 & 1 & i \\
                0 & 0 & i & 1
            \end{pmatrix}_{\mathrm{ct}} \otimes \hat{I}_{\mathrm{s}},
        \end{align}
        where the subscript $\mathrm{ct}$ denotes the control--target subspace and $\hat{I}_{\mathrm{s}}$ is the identity operator on the spectator qubit.
        
        Let $P_{\mathrm{comp}} = \sum_{\bm n\in\mathcal{C}(n)} \ket{\bm n}\!\bra{\bm n}$ be the projector onto the computational subspace.
        The experimentally realized time evolution is a unitary operator $\hat{U}_{\mathrm{exp}}$ on the full Hilbert space, while its restriction to the computational subspace
        \begin{align}
            \tilde U_{\mathrm{exp}} = P_{\mathrm{comp}}\,\hat{U}_{\mathrm{exp}}\,P_{\mathrm{comp}},
        \end{align}
        is not unitary in general due to leakage. We define its matrix elements as
        $u_{\bm m, \bm n} = \bra{\bm n}\hat{U}_{\mathrm{exp}}\ket{\bm m}$.
        
        To clarify the state transition processes that contribute to the gate infidelity, we define the following sets of state transitions:
        \begin{align}
            \mathcal{S}_{ZX} &= \qty{(\bm{m}, \bm{n}) \in \mathcal{C}(n) \times \mathcal{C}(n)~:~(\hat{U}_{ZX})_{\bm m, \bm n} \neq 0}, \\
            \mathcal{S}^{\perp}_{ZX} &= (\mathcal{C}(n) \times \mathcal{C}(n)) \setminus \mathcal{S}_{ZX}, \\
            \mathcal{S}_{\mathrm{leak}} &= \mathcal{C}(n) \times \overline{\mathcal{C}}(n),
        \end{align}
        where we denote $(\hat{U}_{ZX})_{\bm{m, n}} = \bra{\bm{n}}\hat{U}_{ZX}\ket{\bm{m}}$. We note that the matrix elements of $\hat{U}_{ZX}$ have $\abs{\mathcal{S}_{ZX}}=2^{n+1}$ non-zero elements.
    
    \subsection{Infidelity bound}
        The average gate fidelity $F_{\mathrm{avg}}$ of this CR-gate operation is given by~\cite{malekakhlagh2020first}
        \begin{align}
            F_{\mathrm{avg}} = \frac{ \abs{\Tr\!\qty[\hat{U}_{ZX}^{\dagger}\,\tilde U_{\mathrm{exp}}]}^{2}
            \;+\; \Tr\!\qty[\tilde U_{\mathrm{exp}}^{\dagger}\tilde U_{\mathrm{exp}}]}{2^n(2^n+1)}, \label{eq:average_gate_fidelity}
        \end{align}
        where $n$ is the number of qubits defining the computational subspace.

        Consider the first term in Eq.~\eqref{eq:average_gate_fidelity}. Because $\hat{U}_{ZX}$ has non-zero elements only in $\mathcal{S}_{ZX}$, 
        \begin{align}
            \Tr(\hat{U}_{ZX}^{\dagger}\tilde U_{\mathrm{exp}}) = \sum_{(\bm m, \bm n) \in \mathcal{S}_{ZX}} (\hat{U}_{ZX})^*_{\bm{m, n}} u_{\bm{m, n}}.
        \end{align}

        Using the fact that $\abs{(\hat{U}_{ZX})_{\bm m, \bm n}} = 1/\sqrt{2}$ for all transition processes in $\mathcal{S}_{ZX}$, we can bound the first term as
        \begin{align}
            \abs{\Tr(\hat{U}_{ZX}^{\dagger}\tilde U_{\mathrm{exp}})}^{2}
            &\leq \frac{1}{2} \qty(\sum_{(\bm{m}, \bm{n}) \in \mathcal{S}_{ZX}} \abs{u_{\bm m, \bm n}})^2 \\
            &\leq\frac{1}{2} \abs{\mathcal{S}_{ZX}} \sum_{(\bm{m}, \bm{n}) \in \mathcal{S}_{ZX}} \abs{u_{\bm m, \bm n}}^2 \\
            &= 2^n \sum_{(\bm{m}, \bm{n}) \in \mathcal{S}_{ZX}} \abs{u_{\bm m, \bm n}}^2, \label{eq:bound2}
        \end{align}
        where we have used the triangle inequality in the first line and the Cauchy-Schwarz inequality in the second line.

        Applying the unitarity of $\hat{U}_{\mathrm{exp}}$ for each column $\bm n$, $\sum_{\bm m\in\mathcal{F}(n)}\abs{u_{\bm m, \bm n}}^{2}=1$, we can transform
        \begin{align}
            \sum_{(\bm m,\bm n)\in\mathcal{S}_{ZX}} \mkern-25mu \abs{u_{\bm m, \bm n}}^{2}
            = 2^n
            - \mkern-25mu \sum_{(\bm m,\bm n)\in\mathcal{S}_{ZX}^{\perp}} \mkern-25mu\abs{u_{\bm m, \bm n}}^{2}
            - \mkern-25mu \sum_{(\bm m,\bm n)\in\mathcal{S}_{\mathrm{leak}}} \mkern-25mu \abs{u_{\bm m, \bm n}}^{2}.
            \label{eq:partition}
        \end{align}

        The second term in Eq.~\eqref{eq:average_gate_fidelity} is similarly bounded using the unitarity condition of $\hat{U}_{\mathrm{exp}}$ as
        \begin{align}
            \Tr\qty[\tilde U_{\mathrm{exp}}^{\dagger}\tilde U_{\mathrm{exp}}] &= 
            \sum_{\bm n\in\mathcal{C}(n)}\sum_{\bm m\in\mathcal{C}(n)} \abs{u_{\bm m, \bm n}}^{2} \\
            &= \sum_{\bm n\in\mathcal{C}(n)}\qty(1-\sum_{\bm m\in\overline{\mathcal{C}}(n)} \abs{u_{\bm m, \bm n}}^{2}) \\
            &= 2^n - \sum_{(\bm m,\bm n)\in \mathcal{S}_{\mathrm{leak}}}\abs{u_{\bm m, \bm n}}^{2}. \label{eq:bound1}
        \end{align}

        Combining Eqs.~\eqref{eq:bound2}, \eqref{eq:partition}, and \eqref{eq:bound1}, we obtain the following lower bound on the average gate infidelity
        \begin{align}
            \epsilon_{\mathrm{pop}}
            &= 1 - F_{\mathrm{avg}} \\
            & \geq \frac{1}{2^n+1} \mkern-5mu \sum_{(\bm{m}, \bm{n}) \in \mathcal{S}^{\perp}_{ZX}} \mkern-25mu \abs{u_{\bm m, \bm n}}^{2} + \frac{1}{2^n} \mkern-5mu \sum_{(\bm{m}, \bm{n}) \in \mathcal{S}_{\mathrm{leak}}} \mkern-25mu \abs{u_{\bm m, \bm n}}^{2} \\
            & \Rightarrow \frac{1}{2^n+1} \mkern-5mu \sum_{(\bm{m}, \bm{n}) \in \mathcal{S}^{\perp}_{ZX}} \mkern-25mu R_{\bm m, \bm n} + \frac{1}{2^n} \mkern-5mu \sum_{(\bm{m}, \bm{n}) \in \mathcal{S}_{\mathrm{leak}}} \mkern-25mu R_{\bm m, \bm n},
        \end{align}
        where we have replaced transition probabilities with long-time-averaged transition-matrix elements $\abs{u_{\bm m,\bm n}}^{2} \Rightarrow R_{\bm m,\bm n}$ [Eq.~\eqref{eq:lta_transition_probability1}].

        This population-transfer infidelity evaluates the population that leaves the intended transition set $\mathcal{S}_{ZX}$ within the computational subspace as well as population leaking out of the computational subspace. The total infidelity is then obtained by adding the contribution from the residual $ZZ$ interaction discussed in Appendix~\ref{sec:zz_error}.

\section{Details of the analytical yield curve} \label{app:yield_derivation}
    In this section, we present a detailed derivation of the analytical yield curve shown in Fig.~\ref{fig:yield-01pct}(b). 

    For the optimized frequencies $\qty{\omega^{\mathrm{opt}}_i}$, we numerically simulate a chip realization by randomly shifting the optimized frequencies,
    \begin{align}
        \omega^{\mathrm{opt}, *}_i = \omega^{\mathrm{opt}}_i + \Delta_i,
    \end{align}
    where each random qubit-frequency shift $\Delta_{i}$ is drawn independently from a Gaussian distribution $\mathcal{N}(0, \sigma_{\mathrm{f}}^2)$. We declare that a chip realization is collision-free if every standardized margin [Eq.~\eqref{eq:dp_def}] is positive:
    \begin{align}
        \forall p \in \mathcal{P},\ (i,j,k) \in \mathcal{E}_{3}, \quad d_{p}(\bm{\omega}^{\mathrm{opt}, *}_{i,j,k}) > 0,
    \end{align}
    where $\mathcal{P}$ is the set of all collision processes, $\mathcal{E}_{3}$ is the set of three-qubit chains [Eq.~\eqref{eq:maxmin_compact}].

    In practice, the random shift of the $i$th qubit frequency affects multiple standardized margins whose three-qubit chains include the $i$th qubit. For simplicity, here we neglect these correlations and treat the corresponding variations in the standardized margins as independent. Under this approximation, the uncertainty of each standardized margin is $\delta d = \sigma_{\mathrm{f}}$, independent of the process and the three-qubit chain. Therefore, the chip yield is approximately given by
    \begin{align}
        P_{\mathrm{alive}} &\simeq \prod_{p} \prod_{(i,j,k) \in \mathcal{E}_{3}} \mathrm{Prob}\qty[d_{p}(\bm{\omega}^{\mathrm{opt}, *}_{i,j,k}) + \delta d > 0] \\
        &= \prod_{p} \prod_{(i,j,k) \in \mathcal{E}_{3}} \Phi \qty(\frac{d_{p}(\bm{\omega}^{\mathrm{opt}, *}_{i,j,k})}{\sigma_{\mathrm{f}}}). \label{eq:full_yield}
    \end{align}

    The product in Eq.~\eqref{eq:full_yield} is taken over all collision processes and three-qubit chains. Here, we make the further simplifying assumption that the yield is largely determined by the standardized margins near the optimized objective value $z^{\mathrm{opt}}$. We refer to such margins as the active standardized margins. We define a standardized margin as active if the standardized margin before applying the qubit-frequency spread satisfies
    \begin{align}
        d_{p}(\bm{\omega}^{\mathrm{opt}}_{i,j,k}) \leq 1.01 \cdot z^{\mathrm{opt}},
    \end{align}
    where the coefficient $1.01$ is chosen to include the active standardized margins slightly above the threshold. For non-active standardized margins, $d_{p}(\bm{\omega}^{\mathrm{opt}}_{i,j,k}) > 1.01 z^{\mathrm{opt}}$, we assume $\Phi(d_{p}(\bm{\omega}^{\mathrm{opt}, *}_{i,j,k})/\sigma_{\mathrm{f}}) \approx 1$. Therefore, we can approximate Eq.~\eqref{eq:full_yield} by considering only the active standardized margins, which yields
    \begin{align}
        P_{\mathrm{alive}} \simeq \Phi\qty(\frac{z^{\mathrm{opt}}}{\sigma_{\mathrm{f}}})^{N_{\mathrm{active}}}, \label{eq:yield_active}
    \end{align}
    where $N_{\mathrm{active}}$ is the number of active standardized margins.

    Table~\ref{tab:full_yield} shows the number of active and total standardized margins for the 1024-qubit lattices with optimized frequencies at the $0.1\%$ infidelity threshold. We find that $N_{\mathrm{active}}$ remains on the order of the total number of qubits across the different connection topologies and operating regimes considered in this work. This suggests that the \sii{mixed-integer} programming optimizer increases the objective value $z$ until near-critical constraints are distributed broadly over the entire unit cell that constitutes the 1024-qubit lattice. Consequently, the yield curves shown in Fig.~\ref{fig:yield-01pct}(b) are well approximated by the analytical expression in Eq.~\eqref{eq:yield_active} with $N_{\mathrm{active}}$ on the order of the total qubit number.

    \begin{table}
        \sii{
        \centering
        \caption{Number of total standardized margins, $N_{\mathrm{margin}}$, and active standardized margins, $N_{\mathrm{active}}$, for the optimized 1024-qubit lattices at the $0.1\%$ infidelity threshold.}
        \label{tab:full_yield}
        \begin{tabular}{llrr}
        \hline \hline
        Regime & Lattice & $N_{\mathrm{margin}}$ & $N_{\mathrm{active}}$  \\
        \hline
        Straddling & Square & 45714 & 1144 \\
        Straddling & Hexagonal & 29234 & 1096 \\
        Straddling & Heavy-hexagonal & 21576 & 1200 \\
        Far-detuned & Square & 57050 & 1265 \\
        Far-detuned & Hexagonal & 39183 & 2106 \\
        Far-detuned & Heavy-hexagonal & 27504 & 1744 \\
        \hline \hline
        \end{tabular}
        }
    \end{table}

\section{Straddling-regime collision map} \label{app:straddling_collision_map}
\begin{figure*}
        \centering
        \includegraphics[]{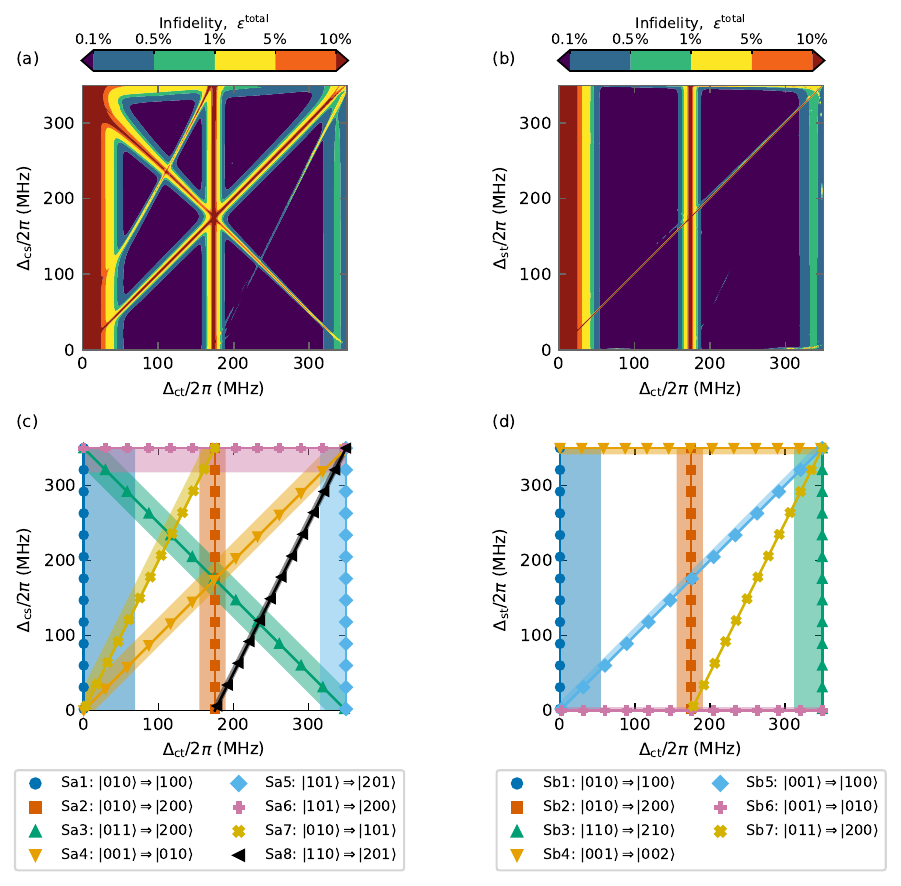}
        \caption{
        Frequency-collision landscapes and frequency-collision maps for three-qubit chains in the straddling regime at the $0.1\%$ infidelity threshold.
        (a) and (b) Frequency-collision landscapes for the target--control--spectator (t-c-s) and control--target--spectator (c-t-s) topologies, respectively. Bright regions indicate parameter regions where the total infidelity $\epsilon_{\mathrm{total}}$ exceeds $0.1\%$. Qubit frequencies can be assigned in the dark regions that avoid all collisions.
        (c) and (d) Frequency-collision maps constructed from the frequency-collision landscapes in (a) and (b) at the $0.1\%$ infidelity threshold for the t-c-s and c-t-s topologies, respectively. The highlighted regions indicate parameter regions where the infidelity exceeds $0.1\%$, and the solid center lines indicate the on-resonance conditions, $\pi_{\bm m,\bm n}(\bm{\omega})=0$.
        }
        \label{fig:straddling_collision_map}
    \end{figure*}

    We show the frequency-collision landscapes and frequency-collision maps for the straddling regime in Fig.~\ref{fig:straddling_collision_map}. The parameter sweeps are performed using the same method as described in the main text except for the qubit--qubit detuning conditions.

\section{Full collision table} \label{app:full-threshold-table}
    We present the full collision conditions in Tables~\ref{tab:riken_thresholds}(a)--(d) for the straddling-regime and far-detuned-regime designs. Each table lists the collision processes, their corresponding effective-detuning expressions, and the lower and upper bounds of the collision windows at the 1\%, 0.5\%, and 0.1\% total-infidelity thresholds. A dash (--) indicates that the corresponding bound is not defined on that side, either because the infidelity peak never exceeds the threshold within the scanned range or because the collision window is intrinsically one-sided.

\section{Collision-free chip size with current qubit-frequency spread}
    \begin{figure*}
        \centering
        \includegraphics[]{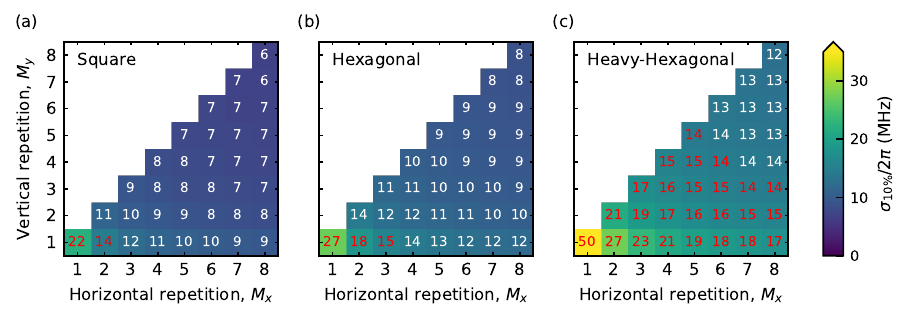}
        \caption{Qubit-frequency spread required to achieve a 10\% collision-free chip yield at the infidelity threshold of $\epsilon_{\mathrm{total}} = 0.1\%$. The chip is constructed by repeating a $4\times4$ unit cell (Fig.~\ref{fig:lattices}) $M_x$ times horizontally and $M_y$ times vertically. (a)--(c) show the required qubit-frequency spread $\sigma_{10\%}$ for square, hexagonal, and heavy-hexagonal lattices, respectively, with optimized frequencies in the far-detuned regime. Red (white) numbers indicate that the requirement is above (below) the minimum state-of-the-art post-fabrication qubit-frequency spread, $\sigma_\mathrm{f}/2\pi = 14.5~\mathrm{MHz}$~\cite{hertzberg2021laser}.}
        \label{fig:yield_heatmap}
    \end{figure*}
    To assess the maximum collision-free QPU size achievable with the current state-of-the-art post-fabrication tuning precision~\cite{hertzberg2021laser}, we perform Monte Carlo simulations using optimized frequency allocations for the far-detuned CR design at the 0.1\% infidelity threshold while varying the chip size and lattice topology.
    For a chip formed by tiling a $4\times4$ unit cell $M_x$ times horizontally and $M_y$ times vertically, we sweep the qubit-frequency spread $\sigma_{\mathrm{f}}$ and determine $\sigma_{10\%}$. Figure~\ref{fig:yield_heatmap} shows the required qubit-frequency spread $\sigma_{10\%}$ as a function of chip size, where white~(red) entries indicate requirements below (above) the state-of-the-art qubit-frequency spread of $\sigma_{\mathrm{f}}/2\pi \approx 14.5\,\mathrm{MHz}$~\cite{hertzberg2021laser}. 

    \begin{table*}[htbp]
        \sii{
        \centering
        \caption{Collision processes and their effective-detuning windows for the far-detuned and straddling regimes in the t-c-s and c-t-s topologies. For each process, the table lists the lower~($l_p$) and upper~($u_{p}$) bounds of the collision window at total infidelity thresholds of 1\%, 0.5\%, and 0.1\%.}
        \label{tab:riken_thresholds}
        \begin{minipage}{\textwidth}
            \centering
            \text{(a) Far-detuned regime t-c-s topology}\\[0.3em]
            \renewcommand{\arraystretch}{1.15}
            \begin{tabular}{l l l cc cc cc ll}
                \hline \hline
                Process   & Expression & Effective detuning &
                \multicolumn{2}{c@{\hspace{2\tabcolsep}\hspace{1.0em}}}{1\% error~(MHz)} &
                \multicolumn{2}{c@{\hspace{2\tabcolsep}\hspace{1.0em}}}{0.5\% error~(MHz)} &
                \multicolumn{2}{c@{\hspace{2\tabcolsep}\hspace{1.0em}}}{0.1\% error~(MHz)} & 
                \\ 
                \cline{4-5}\cline{6-7}\cline{8-9}\cline{10-11} \noalign{\vskip 1.5pt}
                & & & $l_{p}^{(1\%)}$ & $u_{p}^{(1\%)}$ & $l_{p}^{(0.5\%)}$ & $u_{p}^{(0.5\%)}$ & $l_{p}^{(0.1\%)}$ & $u_{p}^{(0.1\%)}$ \\ 
                \hline
                Fa1 & $\ket{001} \Rightarrow \ket{010}$ & $(- \omega_{\mathrm{s}} + \omega_{\mathrm{t}})/2\pi$ & $-15$ & $15$ & $-19$ & $19$ & $-26$ & $26$ \\    
                Fa2 & $\ket{011} \Rightarrow \ket{002}$ & $(\alpha_{\mathrm{s}} + \omega_{\mathrm{s}} - \omega_{\mathrm{t}})/2\pi$ & $-17$ & $16$ & $-23$ & $20$ & $-34$ & $29$ \\
                Fa3 & $\ket{110} \Rightarrow \ket{020}$ & $(\alpha_{\mathrm{t}} - \omega_{\mathrm{c}} + \omega_{\mathrm{t}})/2\pi$ & -- & $25$ & -- & $31$ & -- & $48$ \\
                Fa4 & $\ket{101} \Rightarrow \ket{002}$ & $(\alpha_{\mathrm{s}} - \omega_{\mathrm{c}} + \omega_{\mathrm{s}})/2\pi$ & -- & $14$ & -- & $14$ & -- & $24$ \\
                Fa5 & $\ket{111} \Rightarrow \ket{022}$ & $(\alpha_{\mathrm{t}} + \alpha_{\mathrm{s}} - \omega_{\mathrm{c}} + \omega_{\mathrm{s}})/2\pi$ & $-19$ & $-9$ & $-20$ & $-10$ & $-21$ & $-6$ \\
                Fa6 & $\ket{011} \Rightarrow \ket{020}$ & $(\alpha_{\mathrm{t}} - \omega_{\mathrm{s}} + \omega_{\mathrm{t}})/2\pi$ & $-3$ & $7$ & $-4$ & $7$ & $-5$ & $7$ \\
                Fa7 & $\ket{010} \Rightarrow \ket{002}$ & $(\alpha_{\mathrm{s}} + 2 \omega_{\mathrm{s}} - 2 \omega_{\mathrm{t}})/2\pi$ & -- & -- & $-2$ & $4$ & $-2$ & $4$ \\
                Fa8 & $\ket{010} \Rightarrow \ket{101}$ & $(\omega_{\mathrm{c}} + \omega_{\mathrm{s}} - 2 \omega_{\mathrm{t}})/2\pi$ & $10$ & $15$ & $10$ & $15$ & $7$ & $16$ \\
                Fa9 & $\ket{011} \Rightarrow \ket{003}$ & $(3 \alpha_{\mathrm{s}} + 2 \omega_{\mathrm{s}} - 2 \omega_{\mathrm{t}})/2\pi$ & $-4$ & $0$ & $-4$ & $0$ & $-6$ & $2$ \\
                Fa10 & $\ket{111} \Rightarrow \ket{020}$ & $(- \alpha_{\mathrm{t}} + \omega_{\mathrm{c}} + \omega_{\mathrm{s}} - 2 \omega_{\mathrm{t}})/2\pi$ & -- & -- & -- & -- & $10$ & $17$ \\
                \hline \hline
            \end{tabular}
        \end{minipage}
        \hfill
        \par\vspace{1em}
	    \begin{minipage}{\textwidth}
	        \centering
	        \text{(b) Far-detuned regime c-t-s topology}\\[0.3em]
	        \renewcommand{\arraystretch}{1.15}
	        \begin{tabular}{l l l cc cc cc ll}
	            \hline \hline
                Process   & Expression & Effective detuning &
                \multicolumn{2}{c@{\hspace{2\tabcolsep}\hspace{1.0em}}}{1\% error~(MHz)} &
                \multicolumn{2}{c@{\hspace{2\tabcolsep}\hspace{1.0em}}}{0.5\% error~(MHz)} &
                \multicolumn{2}{c@{\hspace{2\tabcolsep}\hspace{1.0em}}}{0.1\% error~(MHz)} & 
                \\ 
                \cline{4-5}\cline{6-7}\cline{8-9}\cline{10-11} \noalign{\vskip 1.5pt}
                & & & $l_{p}^{(1\%)}$ & $u_{p}^{(1\%)}$ & $l_{p}^{(0.5\%)}$ & $u_{p}^{(0.5\%)}$ & $l_{p}^{(0.1\%)}$ & $u_{p}^{(0.1\%)}$ \\ 
                \hline
	            Fb1 & $\ket{110} \Rightarrow \ket{020}$ & $(\alpha_{\mathrm{t}} - \omega_{\mathrm{c}} + \omega_{\mathrm{t}})/2\pi$ & -- & $34$ & -- & $30$ & -- & $43$ & & \\
	            Fb2 & $\ket{001} \Rightarrow \ket{100}$ & $(\omega_{\mathrm{c}} - \omega_{\mathrm{s}})/2\pi$ & $-7$ & $16$ & $-12$& $19$ & $-16$& $21$ & & \\
	            Fb3 & $\ket{111} \Rightarrow \ket{030}$ & $(3\alpha_{\mathrm{t}} - \omega_{\mathrm{c}} - \omega_{\mathrm{s}} + 2 \omega_{\mathrm{t}})/2\pi$ & $-14$& $22$ & $-15$& $25$ & $-22$& $33$ & & \\
	            Fb4 & $\ket{101} \Rightarrow \ket{002}$ & $(\alpha_{\mathrm{s}} - \omega_{\mathrm{c}} + \omega_{\mathrm{s}})/2\pi$ & $-16$& $2$ & $-16$& $2$ & $-17$& $5$ & & \\
	            Fb5 & $\ket{011} \Rightarrow \ket{020}$ & $(\alpha_{\mathrm{t}} - \omega_{\mathrm{s}} + \omega_{\mathrm{t}})/2\pi$ & -- & $12$ & -- & $12$ & -- & $16$ & & \\
	            Fb6 & $\ket{101} \Rightarrow \ket{200}$ & $(\alpha_{\mathrm{c}} + \omega_{\mathrm{c}} - \omega_{\mathrm{s}})/2\pi$ & $-5$ & $4$ & $-5$ & $4$ & $-15$& $5$ & & \\
	            Fb7 & $\ket{011} \Rightarrow \ket{030}$ & $(3\alpha _{\mathrm{t}}- \omega_{\mathrm{s}} + \omega_{\mathrm{t}})/2\pi$ & $-14$& -- & $-18$& -- & $-20$& -- & & \\
	            \hline \hline
	        \end{tabular}
	    \end{minipage}
        \hfill
        \par\vspace{1em}
        \begin{minipage}{\textwidth}
            \centering
            \text{(c) Straddling regime t-c-s topology}\\[0.3em]
            \renewcommand{\arraystretch}{1.15}
            \begin{tabular}{l l l cc cc cc ll}
                    \hline \hline
            Process   & Expression & Effective detuning &
            \multicolumn{2}{c@{\hspace{2\tabcolsep}\hspace{1.0em}}}{1\% error~(MHz)} &
            \multicolumn{2}{c@{\hspace{2\tabcolsep}\hspace{1.0em}}}{0.5\% error~(MHz)} &
            \multicolumn{2}{c@{\hspace{2\tabcolsep}\hspace{1.0em}}}{0.1\% error~(MHz)} & 
            \\ 
            \cline{4-5}\cline{6-7}\cline{8-9}\cline{10-11} \noalign{\vskip 1.5pt}
            & & & $l_{p}^{(1\%)}$ & $u_{p}^{(1\%)}$ & $l_{p}^{(0.5\%)}$ & $u_{p}^{(0.5\%)}$ & $l_{p}^{(0.1\%)}$ & $u_{p}^{(0.1\%)}$ \\ 
            \hline
                    Sa1 & $\ket{010} \Rightarrow \ket{100}$ & $(\omega_{\mathrm{c}} - \omega_{\mathrm{t}})/2\pi$ & -- & $54$ & -- & $57$ & -- & $69$ & & \\
                    Sa2 & $\ket{010} \Rightarrow \ket{200}$ & $(\alpha_{\mathrm{c}} + 2 \omega_{\mathrm{c}} - 2 \omega_{\mathrm{t}})/2\pi$ & $-28$ & $22$ & $-29$ & $24$ & $-41$ & $29$ & & \\
                    Sa3 & $\ket{011} \Rightarrow \ket{200}$ & $(\alpha_{\mathrm{c}} + 2 \omega_{\mathrm{c}} - \omega_{\mathrm{s}} - \omega_{\mathrm{t}})/2\pi$ & $-21$ & $17$ & $-24$ & $21$ & $-27$ & $28$ & & \\
                    Sa4 & $\ket{001} \Rightarrow \ket{010}$ & $(- \omega_{\mathrm{s}} + \omega_{\mathrm{t}})/2\pi$ & $-6$ & $14$ & $-9$ & $18$ & $-15$ & $25$ & & \\
                    Sa5 & $\ket{100} \Rightarrow \ket{200}$ & $(\alpha_{\mathrm{c}} + \omega_{\mathrm{c}} - \omega_{\mathrm{t}})/2\pi$ & $-10$ & -- & $-22$ & -- & $-35$ & -- & & \\
                    Sa6 & $\ket{101} \Rightarrow \ket{200}$ & $(\alpha_{\mathrm{c}} + \omega_{\mathrm{c}} - \omega_{\mathrm{s}})/2\pi$ & $-16$ & -- & $-18$ & -- & $-33$ & -- & & \\
                    Sa7 & $\ket{010} \Rightarrow \ket{101}$ & $(\omega_{\mathrm{c}} + \omega_{\mathrm{s}} - 2 \omega_{\mathrm{t}})/2\pi$ & $-31$ & $-3$ & $-31$ & $-2$ & $-32$ & $3$ & & \\
                    Sa8 & $\ket{011} \Rightarrow \ket{102}$ & $(\alpha_{\mathrm{s}} + \omega_{\mathrm{c}} + \omega_{\mathrm{s}} - 2 \omega_{\mathrm{t}})/2\pi$ & $-3$ & $8$ & $-5$ & $8$ & $-11$ & $8$ & & \\
                \hline \hline 
            \end{tabular}

        \end{minipage}
        \hfill
        \par\vspace{1em}
        \begin{minipage}{\textwidth}
            \centering
            \text{(d) Straddling regime c-t-s topology} \\ [0.3em]
            \renewcommand{\arraystretch}{1.15}
            \begin{tabular}{l l l cc cc cc ll}
                    \hline \hline
                Process   & Expression & Effective detuning &
                \multicolumn{2}{c@{\hspace{2\tabcolsep}\hspace{1.0em}}}{1\% error~(MHz)} &
                \multicolumn{2}{c@{\hspace{2\tabcolsep}\hspace{1.0em}}}{0.5\% error~(MHz)} &
                \multicolumn{2}{c@{\hspace{2\tabcolsep}\hspace{1.0em}}}{0.1\% error~(MHz)} & 
                \\ 
                \cline{4-5}\cline{6-7}\cline{8-9}\cline{10-11} \noalign{\vskip 1.5pt}
                & & & $l_{p}^{(1\%)}$ & $u_{p}^{(1\%)}$ & $l_{p}^{(0.5\%)}$ & $u_{p}^{(0.5\%)}$ & $l_{p}^{(0.1\%)}$ & $u_{p}^{(0.1\%)}$ \\
                \hline
                    Sb1 & $\ket{010} \Rightarrow \ket{100}$ & $(\omega_{\mathrm{c}} - \omega_{\mathrm{t}})/2\pi$ & -- & $42$ & -- & $46$ & -- & $55$ & & \\
                    Sb2 & $\ket{010} \Rightarrow \ket{200}$ & $(\alpha_{\mathrm{c}} + 2 \omega_{\mathrm{c}} - 2 \omega_{\mathrm{t}})/2\pi$ & $-21$ & $15$ & $-25$ & $19$ & $-39$ & $31$ & & \\
                    Sb3 & $\ket{100} \Rightarrow \ket{200}$ & $(\alpha_{\mathrm{c}} + \omega_{\mathrm{c}} - \omega_{\mathrm{t}})/2\pi$ & $-12$ & -- & $-24$ & -- & $-38$ & -- & & \\
                    Sb4 & $\ket{001} \Rightarrow \ket{002}$ & $(\alpha_{\mathrm{s}} + \omega_{\mathrm{s}} - \omega_{\mathrm{t}})/2\pi$ & $-9$ & -- & $-9$ & -- & $-9$ & -- & & \\
                    Sb5 & $\ket{001} \Rightarrow \ket{100}$ & $(\omega_{\mathrm{c}} - \omega_{\mathrm{s}})/2\pi$ & $-7$ & $0$ & $-9$ & $0$ & $-12$ & $0$ & & \\
                    Sb6 & $\ket{001} \Rightarrow \ket{010}$ & $(- \omega_{\mathrm{s}} + \omega_{\mathrm{t}})/2\pi$ & $-5$ & -- & $-5$ & -- & $-5$ & -- & & \\
                    Sb7 & $\ket{011} \Rightarrow \ket{200}$ & $(\alpha_{\mathrm{c}} + 2 \omega_{\mathrm{c}} - \omega_{\mathrm{s}} - \omega_{\mathrm{t}})/2\pi$ & -- & -- & $-3$ & $1$ & $-3$ & $0$ & & \\
                \hline \hline
            \end{tabular}
        \end{minipage}
        }
    \end{table*}

    \begin{figure*}[htbp]
        \centering
        \includegraphics[]{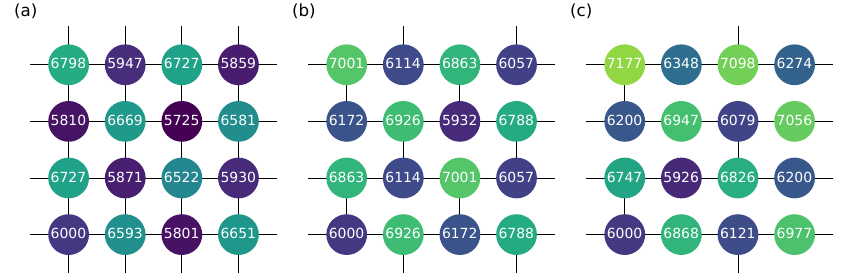}
        \si{
            \caption{Optimized qubit frequencies $\omega_{i}/2\pi$ (MHz) for the far-detuned regime design in the $4\times4$ unit cells of (a)~the square, (b)~hexagonal, and (c)~heavy-hexagonal lattices (node indices shown in Fig.~\ref{fig:lattices}) at the $0.1\%$ infidelity threshold. We fix $\omega_{1}/2\pi=6000~\mathrm{MHz}$ at the left--bottom corner without loss of generality. All anharmonicities are fixed at $\alpha_i/2\pi=-350~\mathrm{MHz}$.}
            \label{fig:optfreq_fardetuned}
        }
    \end{figure*}
    \begin{figure*}[htbp]
        \centering
        \includegraphics[]{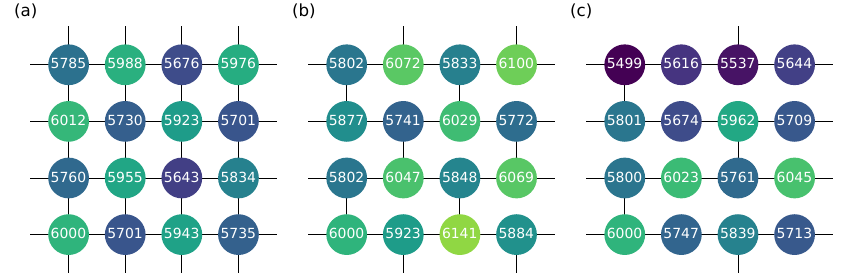}
        \si{
            \caption{Optimized qubit frequencies $\omega_{i}/2\pi$ (MHz) for the straddling regime design in the $4\times4$ unit cells of (a)~the square, (b)~hexagonal, and (c)~heavy-hexagonal lattices at infidelity threshold of $0.1\%$. We fix $\omega_{1}/2\pi=6000~\mathrm{MHz}$ at the left--bottom corner without loss of generality. All anharmonicities are fixed at $\alpha_i/2\pi=-350~\mathrm{MHz}$.}
            \label{fig:optfreq_straddling}
        }
    \end{figure*}

\clearpage
\bibliography{mybib}

\end{document}